\documentclass[
superscriptaddress,
 amsmath,amssymb,
 aps,
prb,two column
]{revtex4-2}

\usepackage{graphicx}
\usepackage{dcolumn}
\usepackage{bm}

\usepackage{graphicx,epsf,bm,bbm,color,amsmath,ulem,amssymb,float, subfigure}

\usepackage{color}
\usepackage{ textcomp }
\usepackage{ dsfont }
\usepackage{indentfirst}
\newcommand{\remove}[1]{}
\def\approxprop{%
  \def\p{%
    \setbox0=\vbox{\hbox{$\propto$}}%
    \ht0=0.6ex \box0 }%
  \def\s{%
    \vbox{\hbox{$\sim$}}%
  }%
  \mathrel{\raisebox{0.7ex}{%
      \mbox{$\underset{\s}{\p}$}%
    }}%
}
\newcommand{\bi}{\begin{itemize}}
\newcommand{\ei}{\end{itemize}}
\newcommand{\be}{\begin{enumerate}}
\newcommand{\ee}{\end{enumerate}}

\newenvironment{dfn}{{\vspace*{1ex} \noindent \bf Definition }}{\vspace*{1ex}}

\newcommand{\nn}{\nonumber}  %

	\newcommand{\beq}{\begin{eqnarray}}
	\newcommand{\eeq}{\end{eqnarray}}

	\renewcommand{\v}{{\bf v}}

\newcommand{\s}{{apple}}

\begin{document}
\title{Enhanced nonlinear Hall effect by Cooper pairs near superconductor criticality}

\author{Zi-Hao Dong}
\affiliation{International Center for Quantum Materials, School of Physics, Peking University, Beijing 100871, China}

\author{Hui Yang}
\affiliation{William H. Miller III Department of Physics and Astronomy, Johns Hopkins University, Baltimore, Maryland, 21218, USA}

\author{Yi Zhang}
\affiliation{International Center for Quantum Materials, School of Physics, Peking University, Beijing 100871, China}
\affiliation{Collaborative Innovation Center of Quantum Matter, Beijing 100871, China
}

\date{\today}

\begin{abstract}
Unlike the linear Hall effect that requires broken time-reversal symmetry, the nonlinear Hall effect may occur in time-reversal symmetric systems as long as there exists a non-zero Berry curvature dipole in the absence of inversion symmetry. Interestingly, the presence of time-reversal symmetry is consistent with and thus allows a direct transition into a superconducting phase. Indeed, superconductivity has been established in various nonlinear Hall materials, such as WTe$_2$ and MoTe$_2$, at sufficiently low temperatures. We find that the nonlinear Hall response should be significantly enhanced near the superconducting criticality, dominated by the Aslamazov-Larkin (AL) contributions augmented by superconducting fluctuations, which we attribute to the Berry curvature dipole and a divergent lifetime $\tau\sim (T-T_c)^{-1}$ of the Cooper pairs, instead of the single electrons. Such a controlled enhancement brings the nonlinear Hall effect into various simple experimental observations and practical applicational potentials.
\end{abstract}

\maketitle

\section{Introduction}

The Hall effect generally requires broken time-reversal symmetry:  the conventional Hall effect \cite{10.2307/2369245} and quantum Hall effect \cite{PhysRevLett.45.494, RevModPhys.58.519,stone1992quantum,cage2012quantum, yoshioka2013quantum} occur in the presence of magnetic fields, and the anomalous Hall effect \cite{hall1881xviii,liu2016quantum,RevModPhys.82.1539} originates from spontaneous time-reversal symmetry breaking \cite{PhysRev.95.1154, PhysRevB.53.7010,PhysRevB.59.14915, PhysRevLett.61.2015}. In a time-reversal invariant system, the linear Hall responses contributed by the integrated Berry curvature $\Omega(k)$ vanishes as $\Omega(-k)=-\Omega(k)$. However, when the inversion symmetry is broken, a nonlinear Hall effect can arise despite a time-reversal symmetric system \cite{PhysRevLett.115.216806, PhysRevLett.121.266601, du2021nonlinear}: in the presence of an electric field, the Berry curvature dipole $\partial\Omega(k)$ leads to an anomalous velocity, which, in turn, generates a Hall current as a second-order effect of the electric field. Such nonlinear Hall effect has been observed in time-reversal symmetric materials\cite{Ma2019,tiwari2021giant,Qin_2021,kumar2021room,he2021quantum,huang2023giant, ye2023control}.

On the other hand, it is possible for a time-reversal symmetric metal to transition into a superconducting state at a lower temperature, where electrons of opposite spin and momentum form Cooper pairs. Therefore, the nonlinear Hall platforms offer a unique opportunity for the direct interplay between Hall physics and superconductivity. For instance, WTe$_2$ becomes superconducting at ~2K \cite{doi:10.1063/5.0021350}.
Semiclassically, the nonlinear Hall effect depends positively on the electron relaxation time\cite{PhysRevLett.115.216806,xiao2019theory}. Interestingly, superconducting fluctuations play vital roles in the transport properties \cite{sctransport1,sctransport2,sctransport3} - they suppress the electron density at the Fermi energy, thus the Drude conductivity, yet more importantly, lead to a divergent Cooper-pair relaxation time $\tau_{\text{GL}} \propto (T-T_c)^{-1}$ in the vicinity of the metal-superconductor transition \cite{galitski2001superconducting, larkin2005theory, RevModPhys.90.015009}. Therefore, an enhanced Hall response is conceivable near the superconducting criticality, but attributed to the physics of the Cooper pairs instead of the electrons. Indeed, enhanced nonlinear Hall response has been observed near the critical point in a trigonal superconductor \cite{itahashi2022giant}. Recently, phenomenological theory unrelated to the Berry curvature dipole has been proposed to describe the AC nonlinear Hall effect and the nonlinear Hall effect under magnetic fields in superconducting systems \cite{sonowal2024nonlinear,daido2024rectification,boev2020photon,boev2022contribution,plastovets2023fluctuation, radkevich2022nonlinear,mironov2024ac}.

Here, after introducing the electron attractive interaction and the superconducting fluctuation propagator, we study the nonlinear Hall effect near the superconducting transition ($T\gtrsim T_c$) by a more careful diagrammatic approach. Without loss of generality, we employ a tilted massive Dirac-fermion model in two dimensions (2D) with time-reversal symmetry yet no inversion symmetry for demonstration, effectively describing the double-layer WTe$_2$. We analytically derive its nonlinear Hall responses' asymptotic expressions, and show that the superconducting fluctuations vitally drive them toward different levels of divergence upon approaching the criticality $T\rightarrow T_c$: the density-of-states (DOS)\cite{ioffe1993effect, gray1993interlayer} and Maki-Thompson (MT) contributions \cite{PhysRevB.1.327,10.1143/PTP.39.897} are subdominant, while the Aslamazov–Larkin (AL) contributions \cite{ASLAMASOV1968238} are primary, which can be precisely attributed to the divergent relaxation time of the Cooper pairs, instead of the single electrons, near superconducting criticality. Further, we analyze such phenomenon's dependence on systematic disorder and anisotropy and demonstrate consistent numerical results from both quantum and semiclassical approaches - the latter with the bosonic Cooper pairs' characteristic charge, lifetime, and Berry curvature dipole. We discuss the various potential experimental and material relevance of such Berry-curvature-dipole candidates in the vicinity of superconducting transition temperatures, which offer a new arena for nontrivial and nonlinear Hall responses and transport.

\section{The nonlinear Hall effect and superconductor fluctuation}

While the integral of the Berry curvature vanishes for a system with time-reversal symmetry yet no inversion symmetry, the Berry curvature dipole $\partial_i \Omega_j$ may survive over the Fermi sea in $d$ dimensions \cite{PhysRevLett.115.216806}:
\beq
D_{ij}=\int_{B.Z.}\frac{{\rm d}^d\bm{k}}{(2\pi)^d} n_F \partial_i \Omega_j,
\eeq
where $n_F=1/[{\exp}(\epsilon_{\bm{k}}/T)+1]$ is the Fermi distribution function, resulting in a nonzero nonlinear Hall contribution. Semiclassically, the applied electric field redistributes the Fermi sea and leads to a net accumulated Berry phase proportional to the Berry curvature dipole, resulting in a Hall response at the next order \cite{PhysRevLett.115.216806}:
\beq
    \chi^{xxy}_0(\omega)&=&-\frac{e^3\tau}{2(1+\mathbbm{i}\omega\tau)}\int_{B.Z.}\frac{{\rm d}^d\bm{k}}{(2\pi)^d}n_F\partial_x \Omega_z \nonumber\\
    &\sim& \frac{\mathbbm{i}e^3}{2\omega}D_{xz},
\label{eq:nqhsc}
\eeq
in the clean limit $\omega \tau \rightarrow \infty$.

\begin{figure}
\centering
\includegraphics[width=8cm]{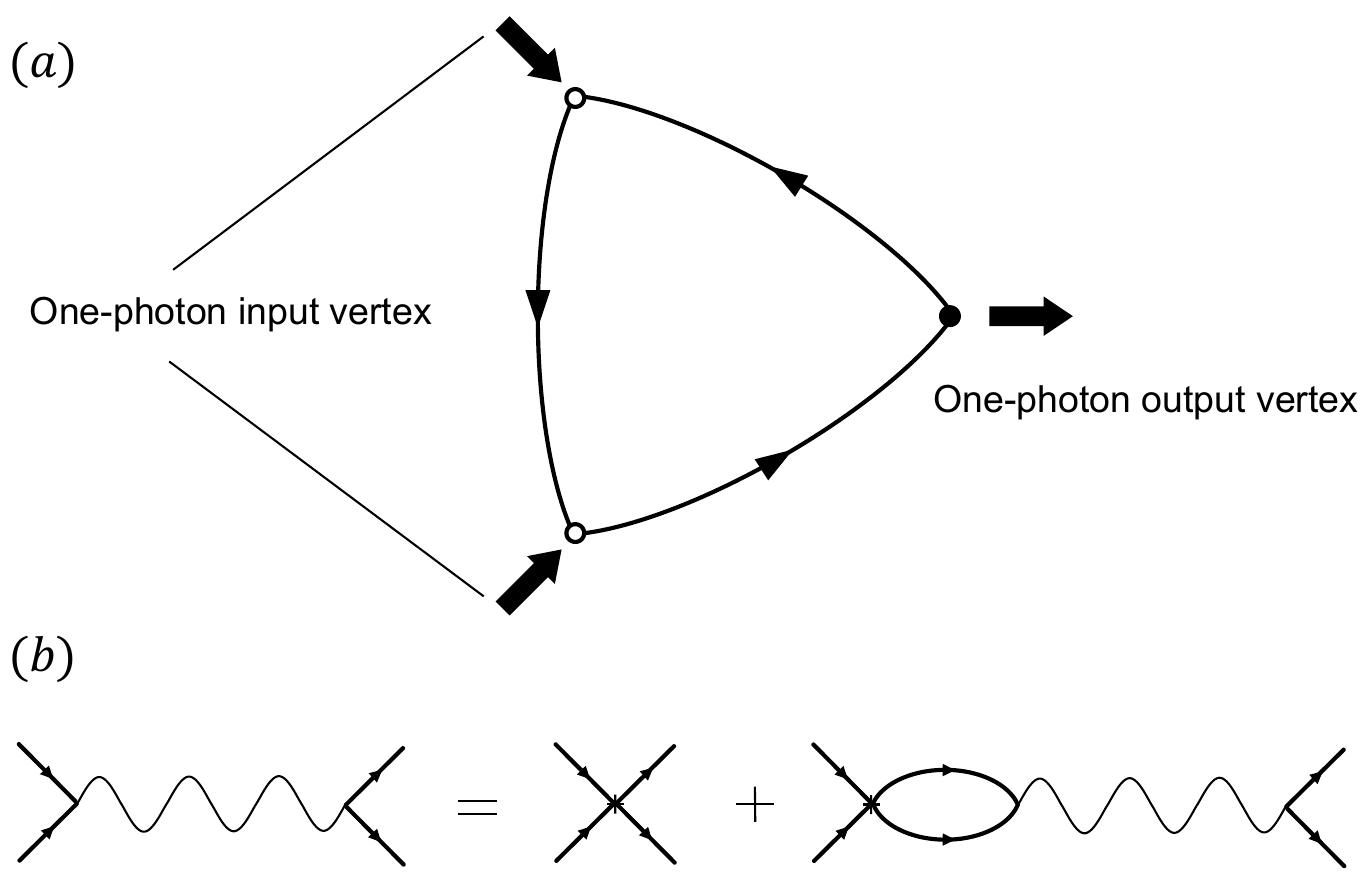}
\caption{(a) We denote the incoming and outgoing photons as empty and solid circles, the fluctuation propagation as the curly line, and the attractive interaction as the (eight-spoked) asterisk, respectively, in the Feynman diagrams of the nonlinear Hall response and (b) the Bethe–Salpeter equation of the fluctuation propagator. }
\label{fig:fd}
\end{figure}

More quantitatively, the nonlinear Hall response corresponds to the Feynman diagram in Fig. \ref{fig:fd}(a), correlating the Berry-curvature-based anomalous velocity and the conventional velocity \cite{parker2019diagrammatic, Du2021}:
\begin{eqnarray}
& &\chi^{xxy}_0(\omega_1,\omega_2)=\frac{Te^3}{\omega_1\omega_2}\int_{B.Z.}\frac{{\rm d}^d\bm{k}}{(2\pi)^d} \\
&\times&\sum_{\epsilon_n} G^a(\epsilon_n, \bm{k}) v_x^{ab} G^b(\epsilon_n+\omega_1, \bm{k}) v_x^{bc} G^c(\epsilon_n+\omega_1 + \omega_2, \bm{k}) v_y^{ca}, \nonumber
\label{eq:normalnqh}
\end{eqnarray}
where $G$'s are the Matsubara Green's functions at imaginary frequencies: $G^a(\epsilon_n, \bm{k})= [\mathbbm{i}\epsilon_n  - E_a(\bm{k})]^{-1}$. $a, b, c$ are band indices to be summed over.

Let's consider a two-band 2D model for concreteness \cite{Du2021}:
\beq
H_{sK}(k)=s t k_x+v( k_x\sigma_x+s k_y\sigma_y)+m \sigma_z, \label{eq:hamk}
\eeq
at the $sK$-valley, where $s=\pm1$. The Pauli matrices $\sigma$'s denote the sublattices, which remain invariant under inversion. Since $\bm{k}\leftrightarrow -\bm{k}$ and $K\leftrightarrow -K$ under time reversal and inversion, the model is time-reversal symmetric [$H_{-K}(\bm{k})=H^*_{K}(-\bm{k})$] yet not inversion symmetric [$H_{-K}(\bm{k})\neq H_K(-\bm{k})$]. Such a (tilted) massive Dirac fermion model offers a minimal effective model for double-layer WTe$_2$, and more generally, carries a representative Berry curvature and Berry curvature dipole distribution for nonlinear Hall responses \cite{PhysRevLett.115.216806}. Without loss of generality, we set $t=0.1$, $v=0.2$, $m=0.3$, and focus on the Fermi energy $E_F=0.3$ (hence a closed electron pocket in the $\bm{k}$ space, see the Appendix) as well as the DC limit at small $\omega\rightarrow 0$. The energy unit is approximately in $O(1)$ eV, as compared with typical electronic dispersion and gap in transition metal dichalcogenide (TMD) materials, e.g., a gap of $\sim 1$eV and a Fermi velocity of $\sim 10^{5}$m/s \cite{PhysRevLett.115.216806}. The details of such settings may influence the responses' proportionalities but not their universal divergent behaviors. Examples and results of other parameters settings are also available in Appendix B. 

By inserting the model's velocity operators and Greens functions into Eq. \ref{eq:hamk}, and summing over the bands with nonzero contributions, we obtain:
\begin{eqnarray}
\chi^{xxy}_0(\omega_1, \omega_2)&=& \frac{\mathbbm{i}e^3(\omega_1+\omega_2)}{4\omega_1\omega_2}\int_{B.Z.}\frac{{\rm d}^2\bm{k}}{(2\pi)^2}  \\
&\times& \left[n_F(E_-)-n_F(E_+)\right]\frac{3mv^4k_x}{2(v^2k^2+m^2)^{\frac{5}{2}}}, \nonumber
\label{eq:modelnhe}
\end{eqnarray}
where we have also employed the analytical continuation $\mathbbm{i}\omega_1\rightarrow\omega_1+\mathbbm{i}\delta, \mathbbm{i}\omega_2\rightarrow\omega_1+\mathbbm{i}\delta$, $\delta \rightarrow 0^+$. $k= \sqrt{k_x^2+k_y^2}$, and $E_{\pm}$ are the dispersion of the upper and lower band, respectively. It is fully consistent with the semiclassical expression in Eq. \ref{eq:nqhsc} with the corresponding Berry curvature dipole:
\beq
\Omega_z=\frac{mv^2}{2(v^2k^2+m^2)^{\frac{3}{2}}}, \partial_x\Omega_z=-\frac{3mv^4k_x}{2(v^2k^2+m^2)^{\frac{5}{2}}},
\eeq
for the model's upper band and its opposite for the upper band. Here, we focus on the scenario where $\omega=\omega_1=\omega_2$ so that the response shows a $2\omega$ frequency; the case of a zero-frequency response for $\omega=\omega_1=-\omega_2$ is similar and yields a proportional susceptibility. As a typical example, we summarize in Fig. \ref{fig:chi}(a) the nonlinear Hall response of the model in Eq. \ref{eq:hamk}, which exhibits a mild temperature dependence.

Another vital transport phenomenon is the advent of superconductivity, for which we include the attractive interaction between the electrons:
\beq
-g\sum_{\bm{k}_1,\bm{k}_2}c^\dag_K(\bm{k}_1)c^\dag_{-K}(-\bm{k}_1)c_{-K}(-\bm{k}_2)c_{K}(\bm{k}_2),
\eeq
which scatters a Cooper pair of electrons with opposite momenta (and valleys) to another pair. $g$ is the strength of the attraction interaction. We also limit the interaction to the upper band since such pairing is only prominent for the degrees of freedom near the Fermi energy. Subsequently, the fluctuation propagator $L(\Omega_k, \bm{q})$ is dictated by the Bethe-Salpeter equation as shown in Fig. \ref{fig:fd}(c):
\beq
L(\Omega_k, \bm{q})=-g+g \cdot \Pi(\Omega_k, q) \cdot L(\Omega_k, \bm{q}), \label{eq:bseq}
\eeq
with the bubble diagram:
\beq
\Pi(\Omega_k, q)&=&T\sum_{\epsilon_n}\int_{B.Z.}\frac{{\rm d}^2\bm{p}}{(2\pi)^2} G(\epsilon_n+\Omega_k, \bm{p}+\bm{q}) G^\prime(-\epsilon_n, -\bm{p}), \nonumber\\
\label{eq:bubble}
\eeq
where $G$ and $G^\prime$ are the Greens functions in the $K$ and $-K$-valley, respectively. Clearly, in the $vq, \Omega_k \ll T$ limit, the fluctuation propagator $L=(\Pi-g^{-1})^{-1}$ diverges as $g \Pi(0,0)=1$, signaling the temperature is lowered towards criticality $T\rightarrow T_c$. Here, we focus on the $T\gtrsim T_c$ scenarios; it leads to a Ginsburg–Landau relaxation time $\tau_{\text{GL}}=\pi/[8(T-T_c)]$ and dominates the systematic transport \cite{larkin2005theory, RevModPhys.90.015009}; still, a long-range superconducting order, which would otherwise expel electric fields and invalidate the nonlinear Hall responses, remains absent.

In turn, we can determine the critical temperature $T_c$ as:
\beq
\Pi(0, 0) = \nu[\phi(\frac{1}{2}+\frac{\omega_D}{2\pi T_c})-\phi(\frac{1}{2})]=g^{-1},
\eeq
where $\phi(x)$ is the derivative of the $\Gamma$ function, $\nu=\int_{B.Z.}\delta(\epsilon-\epsilon_p){{\rm d}^2\bm{p}}/{(2\pi)^2} $ is the density of states (DOS) with respect to the massive Dirac fermion model in Eq. \ref{eq:hamk}, and $\omega_D$ is the Debye frequency, which cuts off of the ``Cooper logarithmic" divergence. Consequently, we obtain:
\beq
T_c=\frac{3.56}{\pi}\omega_De^{-\frac{1}{\nu g}}, \label{eq:Tc}
\eeq
see more details in the Appendix. Here afterward, we set the interaction strength $g=0.23$ and $\omega_D=0.1$ so that we have $T_c=0.001$ as our unit convention. For instance, $T_c\sim 1K$ in monolayer WTe$_2$.

Further, for the nonlinear Hall responses near superconducting criticality, we may expand $\epsilon_{p+q} \approx \epsilon_p+{\bf v_p\cdot q}$ for small $q$, and the fluctuation propagator takes the approximate form:
\beq
L(\Omega_k, \bm{q})=-\frac{\nu^{-1}}{\epsilon+\xi_x^2 q_x^2+\xi_y^2 q_y^2+{\pi\Omega_k}/{8T}+r\sin^2(l_zq_z/2)}, \nonumber\\
\label{eq:svortexq}
\eeq
where we have resumed the Lawrence–Doniach model for the out-of-plane interactions in layered (quasi-2D) superconductors, which can be straightforwardly generalized to 3D systems. Here, $l_z$ is the effective    inter-layer spacing, $r$ is the Lawrence–Doniach anisotropy parameter proportional to the Josephson coupling between the neighboring layers \cite{Lawrence1971THEORYOL}, and:
\beq
\epsilon=\ln(T/T_c) \approx \frac{T-T_c}{T_c}, ~\xi_{i}^2=\frac{7\zeta(3)\langle v_{i}^2 \rangle}{16\pi^2T^2}, ~i=x,y.
\eeq

In the presence of disorder, we impose a mean-free time $\tau$ in the single-particle energy and Greens functions, which in turn renormalize the vertex of the fluctuation propagator:
\beq
\xi_{i}^2(\tau)&=&-\tau^2\langle v_{i}^2 \rangle_{F.S.}\left[ \phi(\frac{1}{2}+\frac{1}{4\pi T\tau})-\phi(\frac{1}{2})-\frac{1}{4\pi T\tau} \phi'(\frac{1}{2}) \right]\nonumber\\
&=&\eta \tau^2\langle v_{i}^2 \rangle_{F.S.},
\label{eq:eta}
\eeq
see detailed calculations in the Appendix.

Throughout our study, we focus on the linear fluctuation regime, which is most valid above a characteristic temperature $\epsilon > Gi$, below which the nonlinear superconducting fluctuations become predominant, and the ladder approximation behind the Bethe–Salpeter equation fails. Here, $Gi$ is the Ginzburg-Levanyuk number, originally defined based on the specific heat:
\begin{eqnarray}
    Gi \simeq  \frac{T_c}{E_F} =3\times10^{-3}.
\label{eq:gi}
\end{eqnarray}
in the 2D clean limit \cite{levanyuk1959contribution}.

\section{Nonlinear Hall response near the metal-superconductor transition}

\begin{figure}
\centering
\includegraphics[width=8cm]{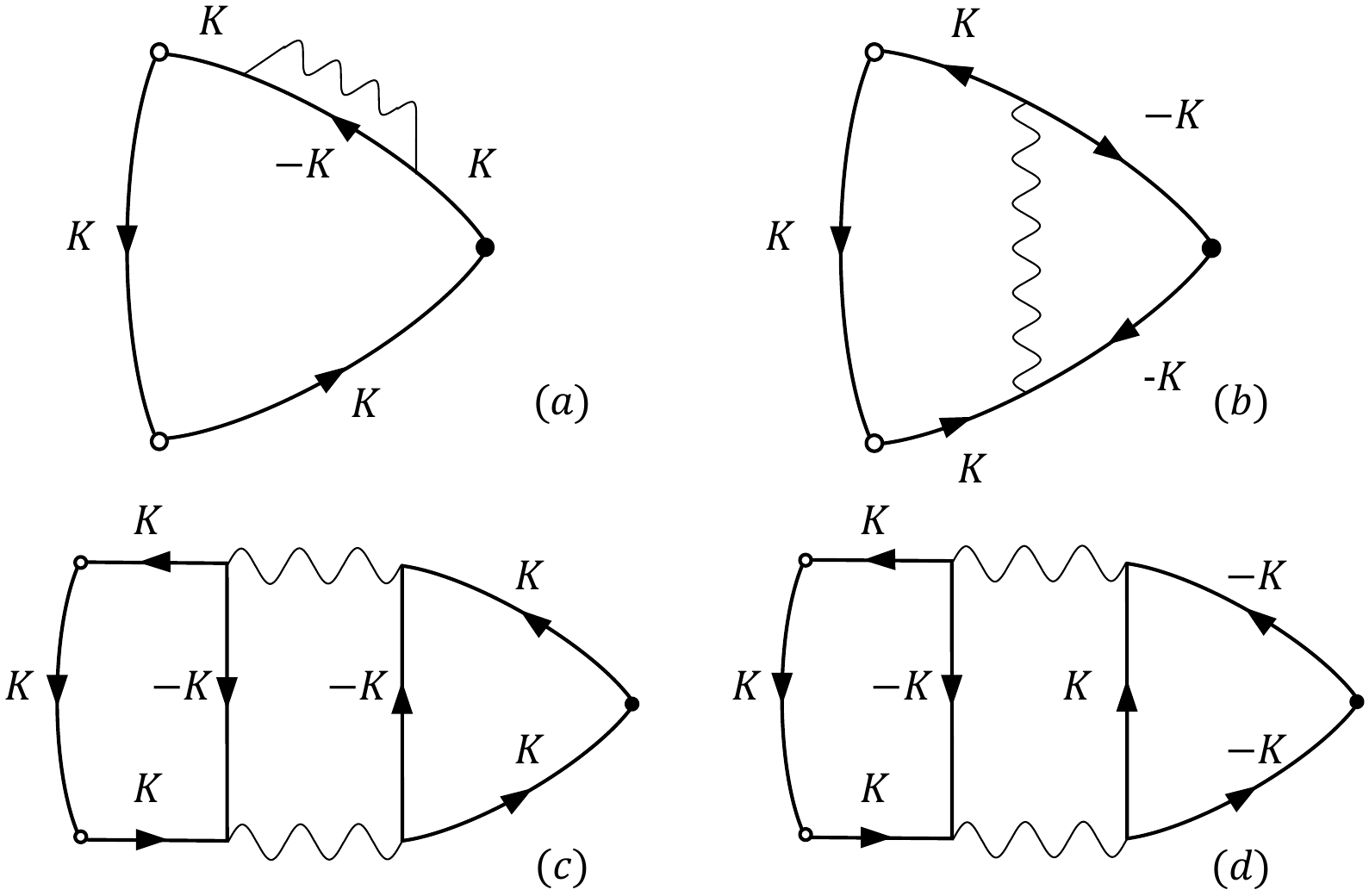}
\caption{We consider the Cooper-pair fluctuations' contributions to the nonlinear Hall response: (a) the contributions from the electron DOS fluctuations due to electrons' participation in Cooper pairs; (b) the MT contributions from the quantum transport process where an electron (hole) is scattered into a Cooper pair and a hole (electron); (c) and (d) the AL contributions from the transport of Cooper pairs. } \label{fig:fd2}
\end{figure}

In addition to the process in Fig. \ref{fig:fd}(a), the nonlinear Hall responses receive vital contributions from the following processes $Q^{xxy}(\omega_1,\omega_2)$ in Fig. \ref{fig:fd2}:
\begin{eqnarray}
    \chi^{xxy}(\omega_1,\omega_2)=-\frac{1}{\omega_1\omega_2}Q^{xxy}(\omega_1,\omega_2),
\end{eqnarray}
given the dominant Cooper-pair fluctuations close to the metal-superconductor transition. Again, we focus on the $\omega=\omega_1=\omega_2$ scenarios, which are straightforwardly generalizable and qualitatively identical to the $\omega=\omega_1=-\omega_2$ cases. Through controlled approximations, we derive analytical expressions of these contributions, which allows us to determine their enhancement and divergence, as well as reliance on systematic properties such as dimensionality and disorders. As we focus on the DC limit without dissipations or dynamical effects, the nonlinear Hall transport is governed exclusively by the real components of the conductivity tensor, isolated from any imaginary contributions.

(1) The process in Fig. \ref{fig:fd2}(a) is the contributions from the DOS fluctuations:
\begin{eqnarray}
&&Q_{{\rm DOS}}^{xxy}(\omega_1,\omega_2)= e^3 T^2\sum\limits_{\Omega_k,\epsilon_n} \int \frac{{\rm d}^3\bm{q}}{(2\pi)^3}\int \frac{{\rm d}^2\bm{p}}{(2\pi)^2}G_+^2(\epsilon_n,\bm{p})\nn\\
&&\times v_x^{++}(\bm{p}) G_+(\epsilon_n+\omega_1,\bm{p})v_x^{+-}(\bm{p})
G_-(\epsilon_n+\omega_1+\omega_2,\bm{p}) \nn\\
&&\times v^{-+}_y(\bm{p})G_+'(\Omega_k-\epsilon_n,\bm{q-p}) L(\Omega_k,\bm{q}) \lambda^2(\epsilon_n,-\epsilon_n,\bm{q}), \nonumber\\
\end{eqnarray}
where for $\epsilon_1\epsilon_2<0$:
\begin{eqnarray}
   \lambda(\epsilon_1,\epsilon_2, \bm{q}) =\frac{|\Tilde{\epsilon_1}-\Tilde{\epsilon_1}|}{|\epsilon_1-\epsilon_2|+\tau\sum_i\langle v_i^2 \rangle_{F.S.}q_i^2},
\end{eqnarray}
renormalizes the vertex of the fluctuation propagator in the presence of disorder.

After some derivation (see the Appendix), we obtain the following analytical expression ($\omega=\omega_1=\omega_2$):
\beq
Q_{\text{DOS}}^{xxy}(\omega)&= \frac{\kappa_{\text{DOS}} e^3 \tau^2\omega^2}{4\pi\xi_x\xi_yl_z}\ln \left[\frac{(\epsilon+r)\left(1-\sqrt{\frac{\epsilon}{\epsilon+r}}\right)^2}{(1+\epsilon+r)\left(1-\sqrt{\frac{1+\epsilon}{1+\epsilon+r}}\right)^2}\right], \nonumber \\ \label{eq:qdos}
\eeq
where $\epsilon = \ln(T/T_c)$, $r$ characterizes the inter-layer coupling in quasi-2D as before, and $\kappa_{\text{DOS}}$ is a non-divergent negative constant depending on $T\tau$ and the specific electron model. Meanwhile, we have neglected the Green's functions' dependence on $\Omega_k$, as $\Omega_k=0$ dominates over its range in the vicinity of the superconducting critical point; we have also kept only the leading order terms in $\bm{q}$, primarily from the fluctuation propagator $L$, as they give the most divergence in the integral.

\begin{figure}
\centering
\includegraphics[width=8cm]{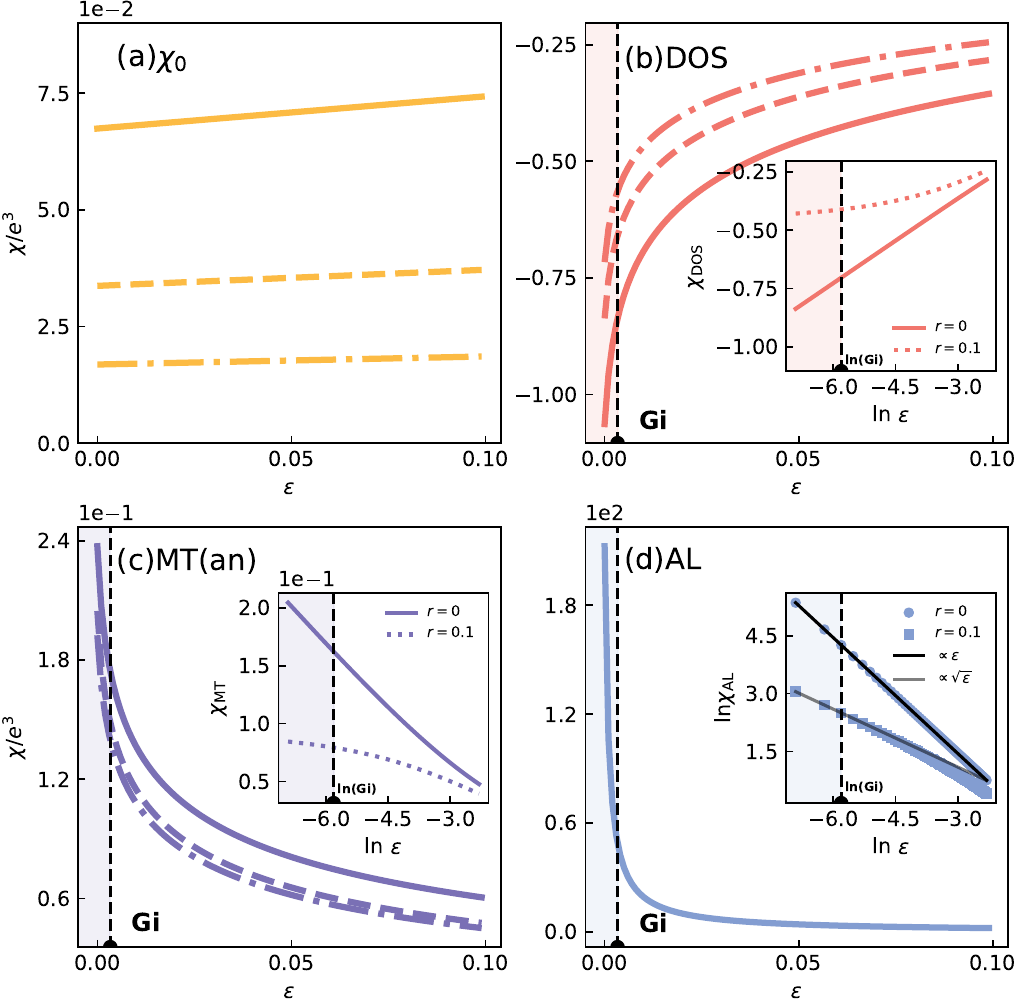}
\caption{The real part of the nonlinear Hall responses $\chi_0$, $\chi_{\text{DOS}}$, $\chi_{\text{MT}}$, and $\chi_{\text{AL}}$ of the model example in Eq. \ref{eq:hamk}, which correspond to the processes in Fig. \ref{fig:fd}(a) and \ref{fig:fd2}(a)-(c), respectively. We only show the anomalous part of $\chi_{\text{MT}}$, whose regular part is on par with $\chi_{\text{DOS}}$ and less divergent. The enhancement and dominance of the AL contributions are prominent. In addition, while a stronger disorder $\tau=0.5T_c^{-1}$ (dot-dashed curves) and
$\tau=T_c^{-1}$ (dashed curves), in comparison to $\tau=2T_c^{-1}$ (solid curves), generally suppresses the other responses, the AL contributions remain mostly resilient. The insets show the nonlinear Hall responses' dependence on the anisotropy parameter $r$, where the AL contributions observe the largest decreases as the dimensionality increases (note the difference in vertical scales). The shaded regions represent non-negligible nonlinear superconducting fluctuations, where $Gi$ denotes the Ginzburg-Levanyuk temperature scale. Here, we set $l_z=1$ and $\gamma_\phi=0.2$. $\epsilon=(T-T_c)/T_c$. } \label{fig:chi}
\end{figure}

The results of the model example in Eq. \ref{eq:hamk} are summarized in Fig. \ref{fig:chi}(b). Note the opposite sign of its contribution with respect to $\chi_0^{xxy}$ in Fig. \ref{fig:chi}(a) - the Cooper pair fluctuations decrease the DOS of single electrons at the Fermi energy, thus the number of direct carriers in charge of the single-electron transport. $Q_{\text{DOS}}^{xxy}$ is not divergent for a finite $r$; in the 2D limit where $r\to 0$, it diverges relatively slowly as $\ln (1+\epsilon^{-1}) \approxprop \ln (T-T_c)$, consistent with the tendency of the electron-density corrections near $T_c$.

(2) The process summarized in Fig. \ref{fig:fd2}(b) is the MT contribution, a purely quantum process where a propagating electron is scattered to a hole and a Cooper pair, and likewise, a propagating hole is scattered to an electron and a Cooper pair:
\begin{eqnarray}
&&Q_{\text{MT}}^{xxy}(\omega_1,\omega_2)= e^3 T^2\sum\limits_{\Omega_k,\epsilon_n} \int \frac{{\rm d}^3\bm{q}}{(2\pi)^3}\int \frac{{\rm d}^2\bm{p}}{(2\pi)^2}G_+(\epsilon_n,\bm{p})\nn\\
&&\times v_x^{++}(\bm{p})^2 G_+(\epsilon_n+\omega_1+\omega_2,\bm{p})L(\Omega_k,\bm{q})\lambda(\epsilon_n,-\epsilon_n,\bm{q})\nn\\
&&\times G_+(\epsilon_n+\omega_1,\bm{p})G_-'(\Omega_k-\epsilon_n-\omega_1-\omega_2,\bm{q-p}) v^{-+}_y(\bm{q-p})\nn\\
&&\times G_+'(\Omega_k-\epsilon_n,\bm{q-p}) \lambda(\epsilon_n+\omega_1+\omega_2,-\epsilon_n-\omega_1-\omega_2,\bm{q}). \nonumber \\
\end{eqnarray}

We may split the MT contributions into a regular part and an anomalous part, whose respective analytical expressions are:
\beq
Q_{\text{MT(reg)}}^{xxy}(\omega)&=& \frac{\kappa^{\text{reg}}_{\text{MT}}e^3 \tau^2 \omega^2}{4\pi\xi_x\xi_yl_z}\ln \left[\frac{(\epsilon+r)\left(1-\sqrt{\frac{\epsilon}{\epsilon+r}}\right)^2}{(1+\epsilon+r)\left(1-\sqrt{\frac{1+\epsilon}{1+\epsilon+r}}\right)^2}\right], \nonumber \\
Q_{\text{MT(an)}}^{xxy}(\omega)&=& \frac{\kappa^{\text{an}}_{\text{MT}} e^3\tau^2 \omega^2}{8\pi\xi_x\xi_yl_z\sqrt{(\epsilon+r-\gamma_\phi)(\epsilon-\gamma_\phi)}} \\\nonumber
&\times&\ln\left[\frac{\sqrt{\frac{1+\epsilon+r}{1+\epsilon}}+\sqrt{\frac{\epsilon+r-\gamma_\phi}{\epsilon-\gamma_\phi}}}{\sqrt{\frac{1+\epsilon+r}{1+\epsilon}}-\sqrt{\frac{\epsilon+r-\gamma_\phi}{\epsilon-\gamma_\phi}}} \frac{\sqrt{\frac{\epsilon+r}{\epsilon}}-\sqrt{\frac{\epsilon+r-\gamma_\phi}{\epsilon-\gamma_\phi}}}{\sqrt{\frac{\epsilon+r}{\epsilon}}+\sqrt{\frac{\epsilon+r-\gamma_\phi}{\epsilon-\gamma_\phi}}}\right],
\eeq
where $\kappa_{\text{MT}}^{\text{an/reg}}$ are model-dependent non-divergent parameters similar to $\kappa_{\text{DOS}}$. $\gamma_\phi={2\tau\eta}/{\tau_\phi}$, $\eta$ is from Eq. \ref{eq:eta}, and $\tau_\phi$ is the phase-breaking time that may originate from the Coulomb interaction, the phonon effect, the superconducting fluctuation, and so on, and cuts off the $\bm{q}$-integration to avoid infrared divergence\cite{PhysRevB.1.327, altshuler1982effects,brenig1985inelastic}. For simplicity, we take $\gamma_\phi$ as a phenomenological constant in our framework. Meanwhile, like in the DOS contributions, we have neglected the Green's functions' $\Omega_k$ dependence; on the other hand, unlike the DOS contributions, we have neglected only the Green's functions' $\bm{q}$ dependence, and kept both $L$'s and $\lambda$'s $\bm{q}$ dependence, as the latter gives rise to the MT contributions' anomalous part, once we sum over $\epsilon_n$ in the range of $[-2\omega,0]$.

The MT contributions of the model example in Eq. \ref{eq:hamk} are summarized in Fig. \ref{fig:chi}(c). We note that the anomalous part always dominates over the regular part as $T\rightarrow T_c$. The regular part $Q_{\text{MT(reg)}}^{xxy}$ displays the same divergent behavior as the DOS contribution in Eq. \ref{eq:qdos}: as $\propto \ln(T-T_c) $ for $r\to 0$ and non-divergent for a finite $r$. On the other hand, $Q_{\text{MT(an)}}^{xxy}$ can become a leading order contribution for $\gamma_\phi\to 0$, on par with the AL contributions that we will discuss next; in realistic scenarios, e.g., $r\rightarrow 0$ and $\gamma_\phi\gg \epsilon$, however, $Q_{\text{MT(an)}}^{xxy}\sim (\epsilon-\gamma_\phi)^{-1}\ln ({\epsilon}/{\gamma_\phi})$ and still diverges as $\propto \ln(T-T_c) $, thus a sub-leading contribution, near superconducting transitions.

(3) The processes in Fig. \ref{fig:fd2}(c) and Fig. \ref{fig:fd2}(d) are the AL contributions, characterizing the participation of Cooper pairs, formed from electrons, in the transport process. Given the system's time-reversal symmetry, these two diagrams give the same contribution:
\begin{eqnarray}
&Q^{\text{AL(c)}}_{xxy}(\omega_1,\omega_2)= -e^3 T\sum\limits_{\Omega_k} \int \frac{{\rm d}^3\bm{q}}{(2\pi)^3} C_{xx}^{(c)}(\Omega_k,\omega_1,\omega_2,\bm{q}) \nn\\
&\times L(\Omega_k,\bm{q})L(\Omega_k+\omega_1+\omega_2,\bm{q}) B_y(\Omega_k,\omega_1+\omega_2,\bm{q}), \nonumber\\
\end{eqnarray}
where:
\begin{eqnarray}
&&B_y(\Omega_k,\omega,\bm{q}) =T\sum_{\epsilon_n} \lambda(\epsilon_n+\omega,\Omega_k-\epsilon_n,\bm{q})  \nonumber\\
&& \times \lambda(\epsilon_n,\Omega_k-\epsilon_n,\bm{q})\int \frac{{\rm d}^2\bm{p}}{(2\pi)^2} v_y^{++}(\bm{p}) G^K_+(\epsilon_n,\bm{p})\nonumber\\
&& \times G^K_+(\epsilon_n+\omega,\bm{p}) G^{-K}_+(\Omega_k-\epsilon_n,\bm{q-p}), \\
&&C_{xx}^{(c)}(\Omega_k,\omega_1,\omega_2,\bm{q}) =T\sum_{\epsilon_n} \lambda(\epsilon_n+\omega,\Omega_k-\epsilon_n,\bm{q})  \nonumber\\
&&\times \lambda(\epsilon_n,\Omega_k-\epsilon_n,\bm{q})\int \frac{{\rm d}^2\bm{p}}{(2\pi)^2} v_x^{-+}(\bm{p}) G^K_-(\epsilon_n,\bm{p}) G^K_+(\epsilon_n+\omega_1,\bm{p})\nonumber\\
&& \times v_x^{++}(\bm{p}) G^K_+(\epsilon_n+\omega_1+\omega_2,\bm{p}) G^{-K}_+(\Omega_k-\epsilon_n,\bm{q-p}).
\end{eqnarray}
By neglecting all the frequency dependence and expanding to the first order of $\bm{q}$ (the zeroth order vanishes) in both in $B$ and $C$, we arrive at the following analytical outcome:
\beq
Q_{\text{AL}}^{xxy}(\omega)
=\frac{\mathbbm{i}\eta e^3\tau^2\omega^2}{\pi l_z m^2}\frac{\left \langle v_x^{++}(\bm{p}) v_x^{+-}(\bm{p}) v_y^{++}(\bm{p})\right \rangle_{F.S.}}{\sqrt{\epsilon(\epsilon+r)} \xi_x\xi_y}, \nonumber \\ \label{eq:ALcontr}
\eeq
where $\eta$ is in Eq. \ref{eq:eta}. Please refer to the Appendix for further details on derivations.

The results of the model example in Eq. \ref{eq:hamk} are summarized in Fig. \ref{fig:chi}(d), which is the most dominant contribution as we approach the superconducting transition temperature $T\to T_c$, offering an enhanced and even divergent nonlinear Hall response. In the 2D limit $r\to 0$, $Q_{\text{AL}}^{xxy}\sim \epsilon^{-1}\propto (T-T_c)^{-1}$, consistent with a semiclassical picture with $\tau_{\text{GL}}\propto (T-T_c)^{-1}$ for Cooper pairs. For quasi-2D systems with a finite $r\gg \epsilon$, $Q_{\text{AL}}^{xxy}\sim \epsilon^{-1/2}\propto (T-T_c)^{-1/2}$, less divergent than the 2D limit yet still dominates over the rest of the nonlinear Hall contributions. Electron models other than Eq. \ref{eq:hamk} modifies the pre-factors such as $\kappa_{\text{DOS}}$, $\kappa_{\text{MT}}^{\text{reg}}$, $\kappa_{\text{MT}}^{\text{an}}$, $\langle\cdots\rangle_{F.S.}$, etc., but the enhancement and divergence behaviors as $T\to T_c$ remain. We note that all divergences at $T\rightarrow T_c$ are practically cutoff at the Ginzburg-Levanyuk temperature scale in Eq. \ref{eq:gi}. Importantly, since $\xi_i^2\propto \eta\tau^2$, the AL contributions are independent of the single-electron mean-free time $\tau$, consistent with the understanding that electron disorders do not affect Cooper pairs' transport may potentially facilitate efficient nonlinear Hall devices and applications \cite{gao2023quantum,gao2024antiferromagnetic}.

Interestingly, such AL contributions are numerically consistent with a straightforward semiclassical interpretation; however, instead of single electrons as in Eq. \ref{eq:nqhsc}, it is based on the physics of Cooper pairs:
\beq
    \chi_{AL}^{xxy}=-\frac{(2e)^3\tau_{\text{GL}}(r)}{2}D_{xz}^{\text{C.P.}},
    \label{eq:Semi-AL}
\eeq
where $2e$, $D_{xz}^{\text{C.P.}}=2D_{xz}$, and:
\begin{eqnarray}
    \tau_{\text{GL}}(r)=\int_0^{\frac{2\pi}{l_z}}\frac{\epsilon\tau_{\text{GL}} \text{d}q_z}{\epsilon+r\sin^2( l_zq_z/2)}=\frac{\epsilon\tau_{\text{GL}}}{l_z\sqrt{\epsilon(\epsilon+r)}},
\end{eqnarray}
are the charge, the Berry curvature dipole, and the effective lifetime of Cooper pairs, respectively.

\begin{figure}
\centering
\includegraphics[width=8cm]{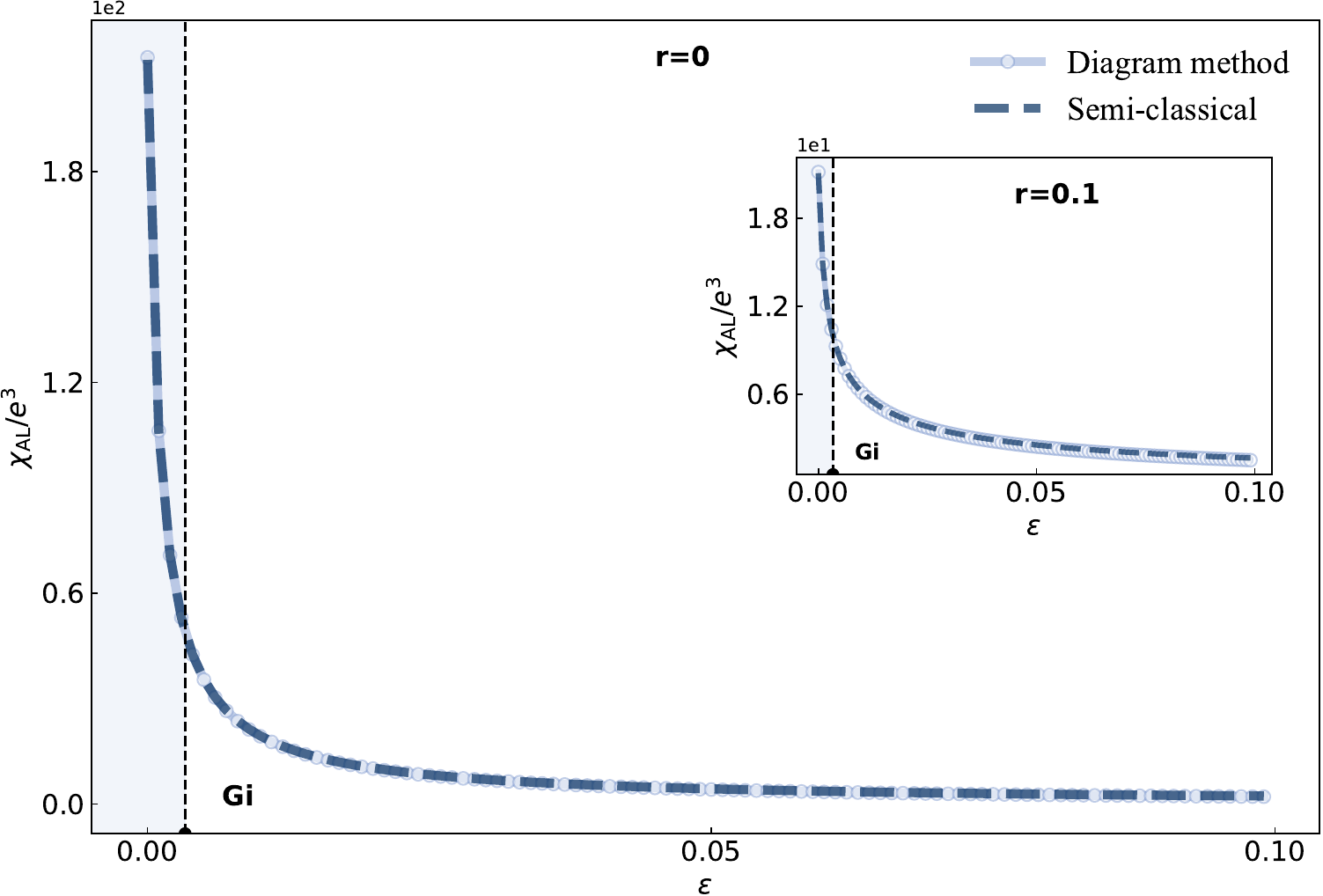}
\caption{The nonlinear Hall response from the semiclassical theory in Eq. \ref{eq:Semi-AL}, based upon Cooper pairs only, compares consistently with the real part of the analytical results of the AL contributions $\chi_{\text{AL}}$ from Eq. \ref{eq:ALcontr}, both in the 2D limit $r\to 0$ and with finite inter-layer coupling $r$ in quasi-2D (inset). $\epsilon=(T-T_c)/T_c$. } \label{fig:semi}
\end{figure}

\begin{figure}
\centering
\includegraphics[width=8cm]{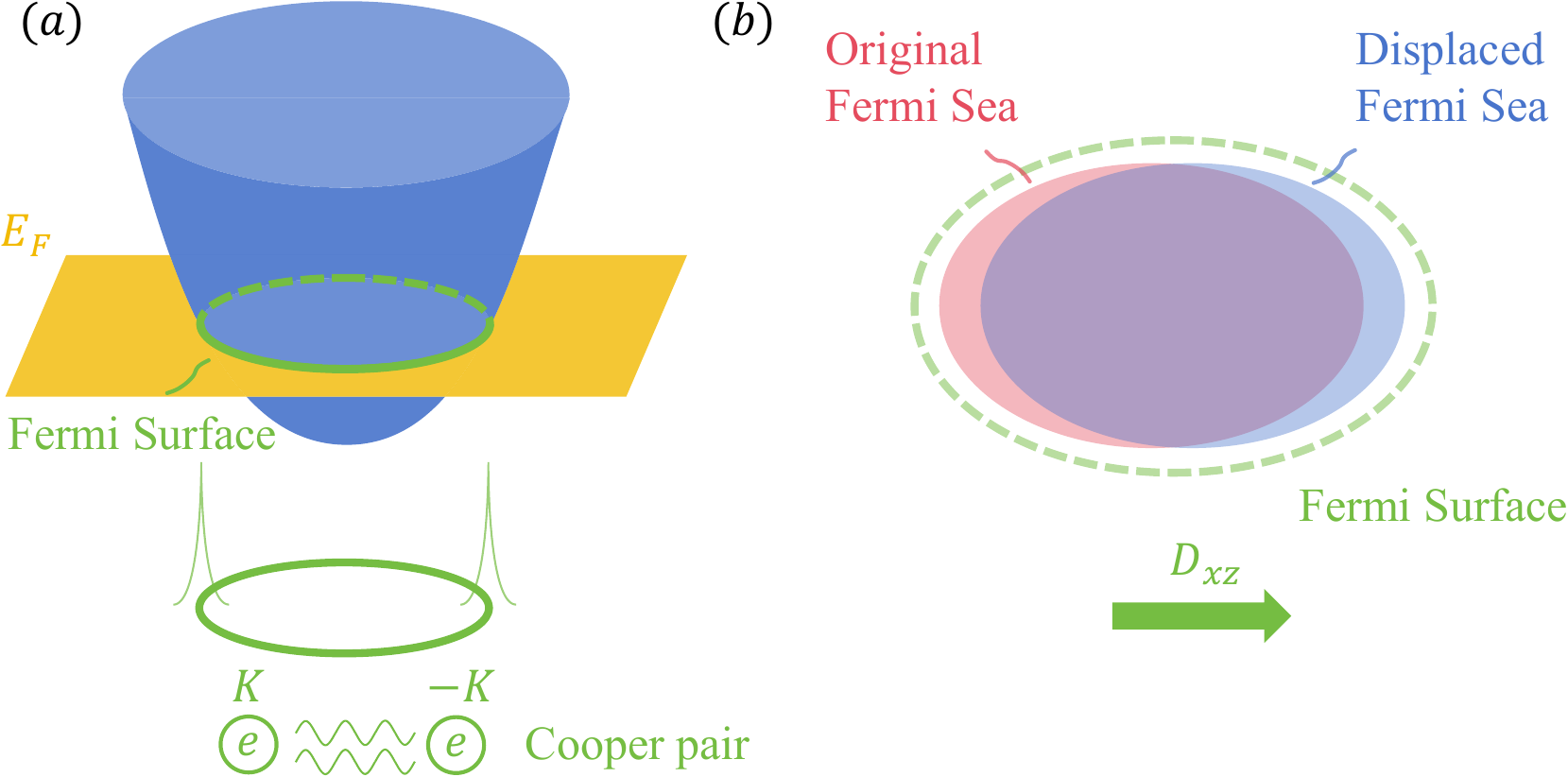}
\caption{Both (a) the Cooper pair fluctuations and (b) the Berry curvature dipole are Fermi surface properties. Given an attractive interaction and temperature down to $T_c$, electrons with opposite momenta near and only near the Fermi surface form Cooper pairs. On the other hand, while the integrated Berry curvature is dependent on the entire Fermi sea, the Berry curvature dipole, effectively the change of the overall integrated Berry curvature under Fermi sea displacement, engages only the Fermi surface, represented by the dashed green circle - contributions from the rest of the Fermi sea (overlapping purple region) cancel. } \label{fig:Cooper}.
\end{figure}

Note that we have presumed $D_{xz}^{\text{C.P.}}=2D_{xz}$ for the Berry curvature dipole of Cooper pairs, doubling that of single electrons. Our theory does not offer a microscopic model for the fluctuating bosonic Cooper pairs, which lacks a well-defined Fermi-surface and thus integrated-Berry-phase description. On the other hand, the single-electron Berry curvature dipole is essentially a Fermi surface property; likewise, despite its bosonic statistics and fluctuation nature, a Cooper pair consists of two electrons near the Fermi surface with opposing momentum and thus identical Berry-curvature-dipole contributions $\partial_i \Omega_j(\boldsymbol{k}) = \partial_i \Omega_j(-\boldsymbol{k})$, given the presenting time-reversal symmetry; see Fig. \ref{fig:Cooper} for illustration. Consequently, we may have a well-defined Berry curvature dipole for Cooper pairs $D_{xz}^{\text{C.P.}}=2D_{xz}$ directly rooting from its single-electron counterpart, and the well consistent numerical results in both the 2D limit $r\to 0$ and quasi-2D systems with a finite out-of-plane effect $r=0$, as shown in Fig. \ref{fig:semi}, offer solid support for such interpretations. Therefore, nonlinear Hall effects near superconducting transitions offer novel experimental and theoretical arenas on Hall and Berry-phase physics of Cooper pairs.

Several approximations, other than Eq. \ref{eq:gi} for the ladder approximation and Gaussian approximation, underline Eq. \ref{eq:Semi-AL}'s semiclassical picture and phenomenological framework of the nonlinear Hall response induced by the Berry curvature dipole of Cooper pairs. First, we prefer a large, well-defined Fermi surface so that the Berry curvature dipole on the Fermi surface contains sufficient information; this suggests that Fermi energy should be sufficiently far away from the band bottom; thus, the Fermi surface dominates the low-energy physics instead of the entire Fermi sea or the thermal broadening. Also, we prefer a smooth Berry curvature distribution, as abrupt variations may introduce numerical singularities and inaccuracies or impacts beyond the Berry curvature dipole that is merely the lowest order description. In Appendix B, we discuss model settings and scenarios where these conditions are approximate, and Eq. \ref{eq:Semi-AL} yields overall semi-quantitative results with observable discrepancy.

\section{Candidate materials}

For the observation of the enhanced nonlinear Hall effect, we recommend searching in the vicinity of superconducting transition temperatures on (quasi-)2D materials with time-reversal symmetry yet no inversion symmetry - preferably materials with higher superconducting $T_c$ and larger Berry curvature dipole, which results in larger Berry curvature dipole of the presenting Cooper pairs. The presence of the nonlinear Hall effect in an ambient environment as single-electron transport is a suitable signature of the latter. Recently, nonlinear Hall effects have been observed in monolayer and multilayer TMD \cite{du2021nonlinear, Ma2019, tiwari2021giant, Qin_2021}, which usually break the inversion symmetry (sometimes by an applied strain) and possess a nontrivial Berry curvature dipole. Interestingly, several monolayer and multilayer TMD materials exhibit superconductivity at low temperatures, e.g., WTe$_2$ exhibits gate-tunable superconductivity at $T_c \sim 1$K \cite{sajadi2018gate,fatemi2018electrically}, multilayer T$_d$-MoTe$_2$ undergoes a metal-superconductor transition at $T_c\sim 0.1$K under ambient pressure \cite{qi2016superconductivity}, and WTe$_2$ shows pressure-induced superconductivity at $T_c\sim 3$K under 2.5GPa \cite{kang2015superconductivity}. Another series of potential candidates involves the Moiré materials of twisted multilayer TMDs where superconductivity has been established - both superconductivity at $T_c\sim 200$mK and nonlinear Hall response have been recently observed in twisted bilayer WSe$_2$ \cite{xia2024superconductivity, huang2023giant}. Such quasi-2D TMD materials offer a realistic platform for the nontrivial interplay between the (enhanced) nonlinear Hall physics and superconductivity.

Our conclusions and material criteria are straightforwardly generalizable to 3D materials. Selected 3D Weyl semimetals may become potential candidates due to their broken inversion symmetry and presenting time-reversal symmetry. For example, the 3D single crystal of T$_d$-TaIrTe$_4$ has exhibited superconductivity at $T_c\sim 2.1$K under 65.7 GPa pressure and nonlinear Hall effect \cite{xing2020surface, cai2019observation, kumar2021room}.

Finally, yet importantly, the surface states of the topological insulators (TIs) also present a promising candidate. The broken inversion symmetry in these materials or by the presence of surfaces themselves and the spin-momentum locking allows a significant surface Berry curvature\cite{he2021quantum}, which may interplay with superconductivity induced through pressure or doping. For instance, Bi$_2$Se$_3$ exhibits superconductivity at $T_c\sim 2-4$K when intercalated with elements such as Cu, Sr, and Nb \cite{hor2010superconductivity,kobayashi2017unusual,liu2015superconductivity,sharma2020superconductivity}.

Besides its considerable enhancement or even divergence, the nonlinear Hall effect near superconductivity has unique and exotic features due to their origin from the Cooper pairs, which may help distinguish it from its pristine single-electron counterpart. Its robustness and resilience against disorder, as apparent in Fig. \ref{fig:chi}(d), is one of its defining features and may harbor useful applications. Also, as a quasi-2D system increases its anisotropic, reduces its thickness, and becomes more 2D in nature ($r$ decreases toward zero), the Cooper pair contribution shows a characteristic increase and eventually a revised scaling in its divergent behaviors in $T-T_c$, the proximity to the transition. Such thickness or layer control is available in (quasi-)2D materials, e.g., with an electric field or gating \cite{2Dthickness1,2Dthickness2,2Dthickness3}. In addition, we have been focusing on the $T\gtrsim T_c$ side of the superconducting transition. In the presence of long-range superconducting order at $T<T_c$, constant or low-frequency electric fields get expelled from the system, and the nonlinear Hall response no longer remains well-defined.

\section{Conclusion}

In conclusion, we study the nonlinear Hall responses in the vicinity of superconducting transitions, such as in selected 3D and 2D materials with time-reversal symmetry and spin-orbit coupling, yet no inversion symmetry. The former validates a direct transition to superconducting phases, while the latter draws the potential for a significant Berry curvature dipole. In particular, we consider the DOS, MT, and AL contributions to the nonlinear Hall effect diagrammatically in the presence of superconducting fluctuations at $T\gtrsim T_c$ and a two-band two-valley (tilted) massive Dirac fermion model in 2D with or without inter-layer coupling and disorder as our demonstration. We attain the analytical expressions of nonlinear Hall contributions asymptotically for $T\rightarrow T_c$, which allows us to determine their respective enhancement and divergence, where the AL contributions dominate.

Our results are also well consistent with a simple semiclassical theory of Cooper pairs, with their characteristic $2e$ charge, divergent lifetime $\tau_{GL}$ at $T\to T_c$, and importantly, Berry curvature dipole $D_{xz}^{\text{C.P.}}=2D_{xz}$ twice the value of its single-electron counterpart. Therefore, the nonlinear Hall responses bring practical meanings of Berry curvature dipole to such bosonic quasiparticles fundamental to superconductivity. We also discuss the enhanced nonlinear Hall responses' robustness against disorder yet suppression by increased dimensionality.

We hope our results and conclusions offer directions for future experiments and applications on candidate materials and inspire further research on the interplay between Cooper-pair transport and Berry-curvature physics. Finally, we note that there are other independent mechanisms, such as skew scattering and side jump, offering enhanced nonlinear Hall responses, e.g., $\propto\tau$ or $\propto\tau^2$, near superconducting transitions \cite{du2021nonlinear}.

\section{Acknowledgment}

We thank Hai-Zhou Lu, An-Yuan Gao, and Zhi-Qiang Gao for the insightful discussions. We also acknowledge support from the National Natural Science Foundation of China (No.12174008 \& No.92270102) and the National Key R\&D Program of China (No.2022YFA1403700).

\bibliography{NonlinearHallEffect.bib}

\begin{widetext}
\appendix

\section{Detailed analytical and numerical results on various Hall responses}

\subsection{The nonlinear Hall effect in the absence of the superconducting fluctuation}

In this appendix, we include further details regarding the derivation of the nonlinear Hall response with respect to the massive Dirac fermion model in Eq. \ref{eq:hamk} in the main text, whose Fermi surface and Berry curvature distribution are illustrated in Fig. \ref{fig:fs}.

\begin{figure}[h]
\centering
\includegraphics[width=15cm]{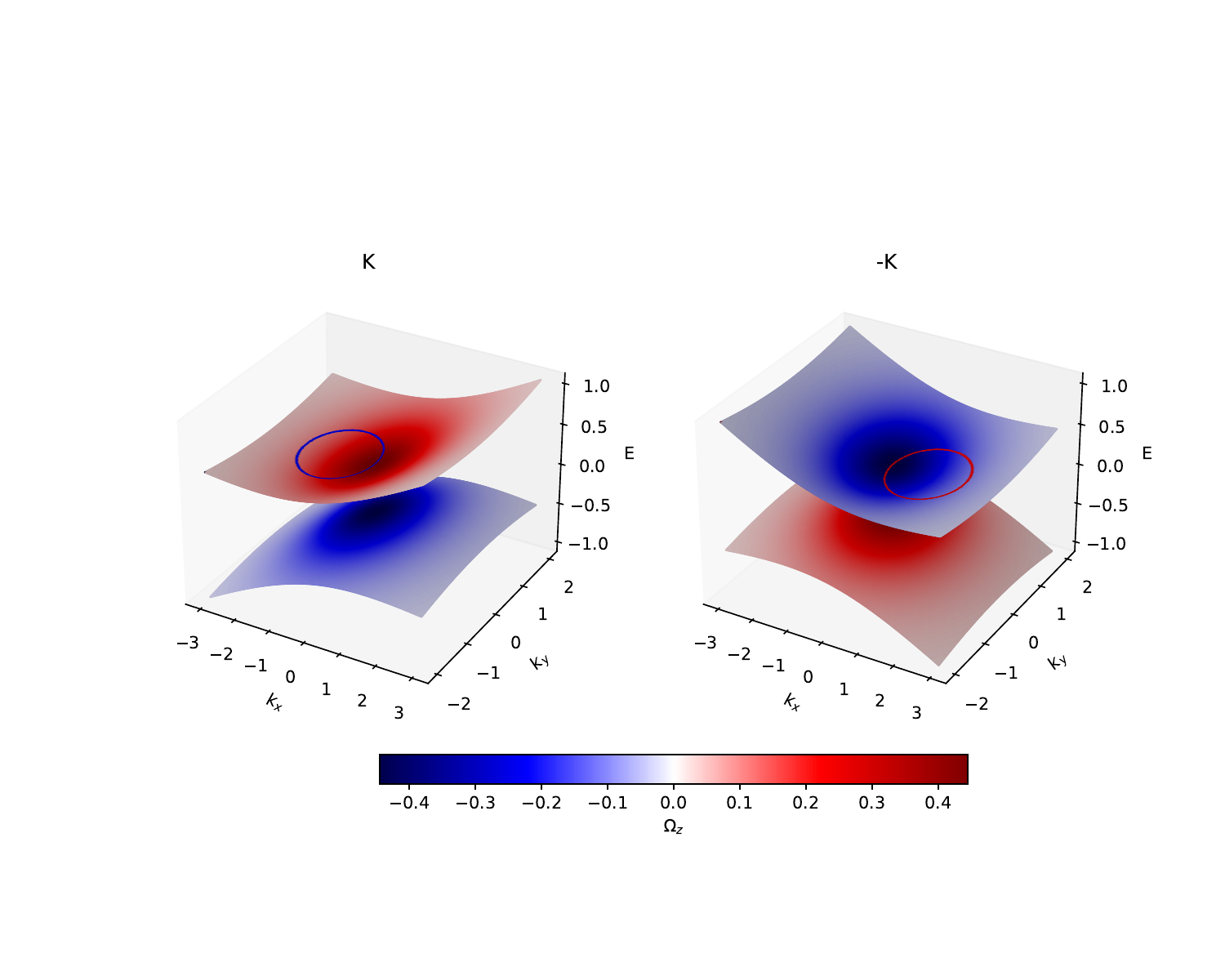}
\caption{The Fermi surface and Berry curvature distribution of the tilted massive Dirac fermion model in Eq. \ref{eq:hamk} in the main text with parameters $t=0.1$, $v=0.2$, $m=0.3$, and $E_F=0.3$. }  \label{fig:fs}
\end{figure}

The corresponding velocity operators are:
\beq
v_x(\bm{k})=\begin{pmatrix}
    st+\frac{v^2k_x}{\sqrt{v^2k^2+m^2}} & \frac{v}{k}(\frac{-mk_x}{\sqrt{v^2k^2+m^2}}-\mathbbm{i}sk_y) \\
    \frac{v}{k}(\frac{-mk_x}{\sqrt{v^2k^2+m^2}}+\mathbbm{i}sk_y) & st-\frac{v^2k_x}{\sqrt{v^2k^2+m^2}}
\end{pmatrix},
v_y(\bm{k})=\begin{pmatrix}
    \frac{v^2k_y}{\sqrt{v^2k^2+m^2}} & \frac{v}{k}(\frac{-mk_y}{\sqrt{v^2k^2+m^2}}+\mathbbm{i}sk_x) \\
    \frac{v}{k}(\frac{-mk_y}{\sqrt{v^2k^2+m^2}}-\mathbbm{i}sk_x) & -\frac{v^2k_y}{\sqrt{v^2k^2+m^2}}
\end{pmatrix}, \nonumber\\
\eeq
which we insert into the pristine nonlinear Hall response in Eq. \ref{eq:normalnqh} in the main text:
\beq
\chi^{xxy}_0(\omega_1, \omega_2)&=&-\frac{e^3}{\omega_1\omega_2}\int_{B.Z.}\frac{{\rm d}^2\bm{k}}{(2\pi)^2}v_x^{ab}v_x^{bc}v_y^{ca}[n_F(E_a)\frac{1}{E_a-E_b+\omega_1}\frac{1}{E_a-E_c+\omega_1+\omega_2}+\nn\\& &
n_F(E_b)\frac{1}{E_b-E_a-\omega_1}\frac{1}{E_b-E_c+\omega_2}+n_F(E_c)\frac{1}{E_c-E_a-\omega_1-\omega_2}\frac{1}{E_c-E_b-\omega_2}] \nn\\
&=&\frac{e^3}{\omega_1\omega_2}\int_{B.Z.}\frac{{\rm d}^2\bm{k}}{(2\pi)^2}\left[n_F(E_-)-n_F(E_+)\right]\mathbbm{i}\frac{mv^4k_x}{v^2k^2+m^2}\frac{3(\omega_1+\omega_2)}{(E_+-E_-)^3},
\eeq
which yields the result in Eq. \ref{eq:modelnhe} in the main text.

\subsection{The fluctuation propagator}

In this Appendix, we evaluate the fluctuation propagator $L^{-1}(\Omega_k, q)=\Pi(\Omega_k, q)-g^{-1}$ via the bubble diagram in Eq. \ref{eq:bubble} in the main text:
\beq
\Pi(\Omega_k, q) &=& T\sum_{\epsilon_n,} \int_{B.Z.}\frac{{\rm d}^2\bm{p}}{(2\pi)^2}G(\epsilon_n+\Omega_k, p+q)G^\prime(-\epsilon_n, -p) \nonumber\\
&=&\int_{B.Z.}\frac{{\rm d}^2\bm{p}}{(2\pi)^2}[n_F(\epsilon_{p+q})\frac{1}{-\epsilon_{p+q}-\epsilon'_{-p}+\Omega_k}-n_F(-\epsilon'_{-p})\frac{1}{-\epsilon_{p+q}-\epsilon'_{-p}+\Omega_k}],
\eeq
where $\epsilon'$ and $G'$ are for the $-K$-valley. For the critical temperature $T_c$, we focus on $\Omega_k=0$, $q=0$:
\beq
\Pi(0, 0)=\int_{B.Z.}\frac{{\rm d}^2\bm{p}}{(2\pi)^2}\left[-n_F(\epsilon_{p})\frac{1}{\epsilon_{p}+\epsilon'_{-p}}+n_F(-\epsilon'_{-p})\frac{1}{\epsilon_{p}+\epsilon'_{-p}}\right]
=-\int_{B.Z.}\frac{{\rm d}^2\bm{p}}{(2\pi)^2}\frac{n_F(\epsilon_{p})-n_F(-\epsilon_{p})}{2\epsilon_{p}},
    \eeq
where we have used the fact that $\epsilon_p=\epsilon'_{-p}$ in the presence of time-reversal symmetry. Further, we simplify the momentum integration as the summation of the residual:
\beq
\Pi(0, 0) =-\sum_{n\geqslant 0}^{\frac{\omega_D}{2\pi T}} \frac{2\nu}{2n+1}=\nu\left[\phi(\frac{1}{2}+\frac{\omega_D}{2\pi T})-\phi(\frac{1}{2})\right],
\eeq
which leads to the expression of $T_c$ - Eq. \ref{eq:Tc} in the main text. $\phi(x)$ is the derivative of the $\Gamma$ function and $\nu=\int_{B.Z.}\delta(\epsilon-\epsilon_p) {{\rm d}^2\bm{p}}/{(2\pi)^2}$ is the density of states.

For the bubble diagram with more general yet small $\Omega_k$ and $q$, we have $\epsilon_{p+q}\approx \epsilon_p+\bm{v_p\cdot q}$, and:
\beq
\Pi(\Omega_k, q)&=& \nu\sum_{n\geqslant 0}\left[\frac{1}{2n+1+\frac{|\Omega_k| }{2\pi T}+\frac{\mathbbm{i} \bm{v_p \cdot q}}{2\pi T}}+\frac{1}{2n+1+\frac{|\Omega_k| }{2\pi T}-\frac{\mathbbm{i} \bm{v_p\cdot q}}{2\pi T}}\right]\nonumber\\
&\approx& \nu\sum_{n\geqslant 0}\left[\frac{2}{2n+1+\frac{|\Omega_k| }{2\pi T}}-\frac{2\left\langle (\bm{v_pq})^2\right\rangle_{F.S.} }{(2n+1+\frac{|\Omega_k| }{2\pi T} )^3(2\pi T)^2}\right],
\eeq
where $\left\langle \right\rangle_{F.S.}=\int \frac{d\theta}{2\pi} $ is the average over the Fermi surface. Inserting $\Pi(\Omega_k, q)$ into the Bethe-Salpeter equation in Eq. \ref{eq:bseq} in the main text, we obtain the fluctuation propagator in Eq. \ref{eq:svortexq} in the main text.

\subsection{The role of the disorder}

We consider a random quenched disorder with uncorrelated on-site potentials as:
\beq
\langle U(r) \rangle =0, \langle U(r)U(r') \rangle = {1}/{2\pi \nu \tau},
\eeq
which affects the single-particle energy and Greens functions:
\beq
\epsilon_n\rightarrow \Tilde{\epsilon_n}=\epsilon_n+\frac{1}{2\tau}{\rm sgn}(\epsilon_n),
\eeq
and the vertex of the fluctuation propagator:
\beq
1\rightarrow \lambda(\epsilon_1,\epsilon_2, \bm{q}) =1+\frac{1}{2\pi \nu \tau}\int \frac{{\rm d}^2\bm{p}}{(2\pi)^2} \lambda(\bm{q},\epsilon_1,\epsilon_2)G(\Tilde{\epsilon_2}, \bm{p}+\bm{q})G^\prime(\Tilde{\epsilon_1}, \bm{-p}).
\eeq

Consequently, we revise the bubble diagram as:
\beq
\Pi(\Omega_k, q) &=& T\sum_{\epsilon_n}\int \frac{{\rm d}^2\bm{p}}{(2\pi)^2}  \lambda(\bm{q},\mathbbm{i}\Tilde{\omega_m}+\Omega_k,-\mathbbm{i}\Tilde{\omega_m}) G(\mathbbm{i}\Tilde{\omega_m}+\Omega_k, \bm{p}+\bm{q}) G^\prime(-\mathbbm{i}\Tilde{\omega_m}, \bm{-p}) \nonumber\\
&\approx& \nu\sum_{n\geqslant 0}\left[\frac{2}{2n+1+\frac{|\Omega_k| }{2\pi T}}-\frac{2\left\langle (\bm{v_pq})^2\right\rangle_{F.S.} }{(2n+1+\frac{|\Omega_k|}{2\pi T} )^2(2n+1+\frac{|\Omega_k|}{2\pi T}+\frac{1}{2\pi T \tau} )(2\pi T)^2}\right],
\eeq
and the retarded fluctuation propagator becomes:
\beq
L^R(\Omega, \bm{q})=-\frac{1}{\nu}\frac{1}{\epsilon+\xi_x^2(\tau) q_x^2+\xi_y^2(\tau) q_y^2-{\mathbbm{i}\pi}\Omega_k/{8T}},
\eeq
where:
\beq
\xi_{i}^2(\tau)&=&-\tau^2\langle v_{i}^2 \rangle_{F.S.}\left[ \phi(\frac{1}{2}+\frac{1}{4\pi T\tau})-\phi(\frac{1}{2})-\frac{1}{4\pi T\tau} \phi'(\frac{1}{2}) \right]\nonumber\\
&=&\eta \tau^2\langle v_{i}^2 \rangle_{F.S.}, \\
\eta &=& -\left[ \phi(\frac{1}{2}+\frac{1}{4\pi T\tau}) - \phi(\frac{1}{2})-\frac{1}{4\pi T\tau}\phi'(\frac{1}{2}) \right],
\eeq
as we have summarized in Eqs. \ref{eq:svortexq}-\ref{eq:eta} in the main text.

\subsection{Nonlinear Hall effect in the presence of superconducting fluctuation}

\begin{figure}[h]
\centering
\includegraphics[width=18cm]{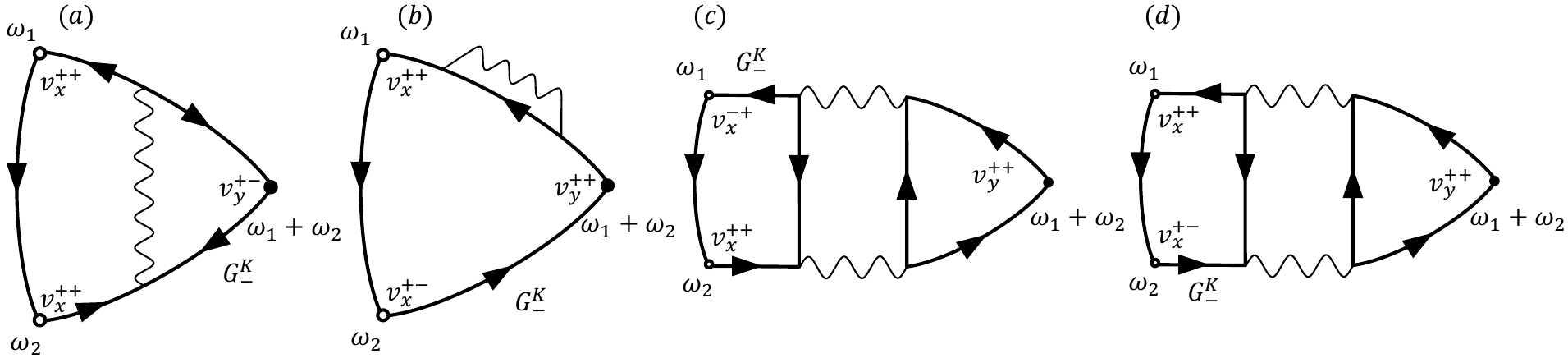}
\caption{Similar to Fig. \ref{fig:fd2} in the main text, we present the Feynman diagrams of the MT contribution, the DOS contribution, and the AL contributions (from left to right) with more detailed notations and labels. }  \label{fig:FDAppendix}
\end{figure}

\subsubsection{The MT contribution}

The contribution to the nonlinear Hall effect from the MT fluctuation, shown in Fig. \ref{fig:FDAppendix}(a), takes the form of:
\beq
Q_{\text{MT}}^{xxy}&= e^3 T\sum_{\Omega_k} \int \frac{{\rm d}^3\bm{q}}{(2\pi)^3} I^{xxy}(\Omega_k,\omega_1,\omega_2,\bm{q}) L(\Omega_k,\bm{q}),
\eeq
where:
\beq
I^{xxy}(\Omega_k=0,\omega_1,\omega_2,\bm{q}) &=& \nu T\sum_{\epsilon_n} \lambda(\epsilon_n,-\epsilon_n,\bm{q}) \lambda(\epsilon_{n+2\omega},-\epsilon_{n+2\omega},\bm{q}) \nonumber\\
&\times&\int_{-\infty}^\infty \frac{d\xi \left\langle {v_x^{++}(\bm{p})v_x^{++}(\bm{p})v_y^{+-}(\bm{q-p})} \right \rangle_{F.S.} }{(\Tilde{\epsilon_n}^2+\xi^2)[\mathbbm{i}(\Tilde{\epsilon_n}+\omega_1+\omega_2)-\xi][\mathbbm{i}(\Tilde{\epsilon_n}+\omega_1)-\xi][\mathbbm{i}(\Tilde{\epsilon_n}+\omega_1+\omega_2)+\xi-2m]},
\eeq
where we have suppressed $\bm{q}$ in the Green's functions. After the integration of $\xi$ and setting $\omega=\omega_1=\omega_2$, we have:
\beq
I^{xxy}(\omega,\bm{q})
&=&\frac{\mathbbm{i}\pi T\nu}{2}\left\langle {v_x^{++}(\bm{p})v_x^{++}(\bm{p})v_y^{+-}(\bm{-p})} \right \rangle_{F.S.} \sum_{\epsilon_n} \lambda(\epsilon_n,-\epsilon_n,\bm{q}) \lambda(\epsilon_{n+2\omega},-\epsilon_{n+2\omega},\bm{q})\nonumber\\
&\times &\left\{  \frac{\theta(\epsilon_n\epsilon_{n+2\omega}){\rm sgn}(\epsilon_n)}{\Tilde{\epsilon_n}(\Tilde{\epsilon_n}+\Tilde{\epsilon_{n+\omega}})(\Tilde{\epsilon_n}+\Tilde{\epsilon_{n+2\omega}})(\omega+\mathbbm{i}m)} + \frac{\theta(-\epsilon_n\epsilon_{n+2\omega})}{\omega+\frac{1}{2\tau}}\left[ \frac{\theta(-\epsilon_n\epsilon_{n+\omega})}{({\epsilon_n}-\frac{1}{2\tau})(\omega+\frac{1}{\tau})({\epsilon_n}+{\epsilon_{n+2\omega}}+2\mathbbm{i}m)}\right.\right. \nonumber\\
 &-&\left.\left.\frac{\theta(\epsilon_n\epsilon_{n+\omega})}{\epsilon_n+\epsilon_{n+2\omega}}\left( \frac{\omega+\frac{1}{2\tau}}{(\omega+\frac{1}{2\tau}+\mathbbm{i}m)(\epsilon_n+\epsilon_{n+\omega}-\frac{1}{\tau})(\epsilon_n-\frac{1}{2\tau})}-\frac{1}{(\omega+\frac{1}{\tau})(\epsilon_{n+2\omega}+\frac{1}{2\tau}+\mathbbm{i}m)} \right) \right] \right\}.
\eeq

We expand $I^{xxy}$ to the $\omega^2$ order and split its contributions into three separate parts:
\beq
I^{xxy}_{1}(\epsilon_n,\omega,\bm{q})&=&\frac{\mathbbm{i}\pi T\nu}{2}\left\langle {v_x^{++}(\bm{p})v_x^{++}(\bm{p})v_y^{+-}(\bm{-p})} \right \rangle_{F.S.} \sum_{\epsilon_n}\frac{\theta(\epsilon_n\epsilon_{n+2\omega}){\rm sgn}(\epsilon_n)\lambda(\epsilon_n,-\epsilon_n,\bm{q})  \lambda(\epsilon_{n+2\omega},-\epsilon_{n+2\omega},\bm{q})}{\Tilde{\epsilon_n}(\Tilde{\epsilon_n}+\Tilde{\epsilon_{n+\omega}})(\Tilde{\epsilon_n}+\Tilde{\epsilon_{n+2\omega}})(\omega+\mathbbm{i}m)}\nonumber\\
&\approx& \frac{\mathbbm{i}\pi T\nu\omega}{m^2}\left\langle {v_x^{++}(\bm{p})v_x^{++}(\bm{p})v_y^{+-}(\bm{-p})} \right \rangle_{F.S.}\sum_{n=0}^\infty \frac{1}{(2\epsilon_n+D_xq_x^2+D_yq_y^2)(2\epsilon_{n+2\omega}+D_xq_x^2+D_yq_y^2)}\nonumber\\
&&\times \left[ \frac{2\epsilon_{n}+4\omega+\frac{1}{\tau}}{(2\epsilon_{n}+2\omega+\frac{1}{\tau})(2\epsilon_{n}+\omega+\frac{1}{\tau})} + \frac{2\epsilon_{n}+\frac{1}{\tau}}{(2\epsilon_{n}+2\omega+\frac{1}{\tau})(2\epsilon_{n}+3\omega+\frac{1}{\tau})}\right]\nonumber\\
&=& \mathbbm{i}\nu\tau^3\left\langle {v_x^{++}(\bm{p})v_x^{++}(\bm{p})v_y^{+-}(\bm{p})} \right \rangle_{F.S.} \frac{\omega^2}{m^2} \left\{ \frac{1}{4\pi T \tau}[\phi'(\frac{1}{2})-\phi'(\frac{1}{2}+\frac{1}{4\pi T \tau})] +\frac{1}{(4\pi T \tau)^2}\phi''(\frac{1}{2}) \right\},
\eeq
\beq
I^{xxy}_{2}(\epsilon_n,\omega,\bm{q})&=&\frac{\mathbbm{i}\pi T\nu}{2}\left\langle {v_x^{++}(\bm{p})v_x^{++}(\bm{p})v_y^{+-}(\bm{-p})} \right \rangle_{F.S.} \sum_{\epsilon_n} \frac{\theta(-\epsilon_n\epsilon_{n+2\omega}) \theta(-\epsilon_n\epsilon_{n+\omega}) \lambda(\epsilon_n,-\epsilon_n,\bm{q})  \lambda(\epsilon_{n+2\omega},-\epsilon_{n+2\omega},\bm{q})} {({\epsilon_n}-\frac{1}{2\tau}) (\omega+\frac{1}{\tau}) (\omega+\frac{1}{2\tau})({\epsilon_n}+{\epsilon_{n+2\omega}}+2\mathbbm{i}m)}\nonumber\\
&=& \frac{\mathbbm{i}\pi T\nu\left\langle {v_x^{++}(\bm{p})v_x^{++}(\bm{p})v_y^{+-}(\bm{-p})} \right \rangle_{F.S.}}{2 (\omega+\frac{1}{\tau}) (\omega+\frac{1}{2\tau})}\sum_{n=0}^{\nu-1} \frac{2\epsilon_{n-2\nu}-\frac{1}{\tau}}{(2\epsilon_n+\bm{Dq}^2)(2\epsilon_{n-2\nu}-\bm{Dq}^2)(2\epsilon_{n-\nu}-2\mathbbm{i}m)}\nonumber\\
&=& -\frac{\mathbbm{i}\nu\tau^3}{4}\left\langle {v_x^{++}(\bm{p})v_x^{++}(\bm{p})v_y^{+-}(\bm{p})} \right \rangle_{F.S.}\frac{\omega^2}{m^2}\left[ \frac{1}{4\pi T \tau}\phi'(\frac{1}{2}) -\frac{1}{(4\pi T \tau)^2}\phi''(\frac{1}{2}) \right],
\eeq
where $D_{i}=\tau\langle v_{i}^2 \rangle_{F.S.}={\xi_{i}^2}/{\eta\tau}$, $i=x,y$. We denote $\bm{Dq}^2\equiv D_xq_x^2+D_yq_y^2$ for simplicity. We further divide the third part:
\beq
I^{xxy}_{3}(\epsilon_n,\omega,\bm{q})&=&-\frac{\mathbbm{i}\pi T\nu}{2}\left\langle {v_x^{++}(\bm{p})v_x^{++}(\bm{p})v_y^{+-}(\bm{-p})} \right \rangle_{F.S.} \sum_{\epsilon_n} \frac{\theta(-\epsilon_n\epsilon_{n+2\omega}) \theta(\epsilon_n\epsilon_{n+\omega})   \lambda(\epsilon_n,-\epsilon_n,\bm{q})  \lambda(\epsilon_{n+2\omega},-\epsilon_{n+2\omega},\bm{q})}{\epsilon_n+\epsilon_{n+2\omega}} \nonumber\\
& &\times \left[ \frac{1}{(\omega+\frac{1}{2\tau}+\mathbbm{i}m)(\epsilon_n+\epsilon_{n+\omega}-\frac{1}{\tau})(\epsilon_n-\frac{1}{2\tau})}-\frac{1}{(\omega+\frac{1}{2\tau})(\omega+\frac{1}{\tau})(\epsilon_{n+2\omega}+\frac{1}{2\tau}+\mathbbm{i}m)} \right],
\eeq
into two terms $I^{xxy}_{3(1)}$ and $I^{xxy}_{3(2)}$:
\beq
I^{xxy}_{3(1)}(\epsilon_n,\omega,\bm{q}) &=& (\cdots)\sum_{\epsilon_n} \frac{\theta(-\epsilon_n\epsilon_{n+2\omega}) \theta(\epsilon_n\epsilon_{n+\omega}) \lambda(\epsilon_n,-\epsilon_n,\bm{q})  \lambda(\epsilon_{n+2\omega},-\epsilon_{n+2\omega},\bm{q})}{(\epsilon_n+\epsilon_{n+2\omega})(\omega+\frac{1}{2\tau}+\mathbbm{i}m)(\epsilon_n+\epsilon_{n+\omega}-\frac{1}{\tau})(\epsilon_n-\frac{1}{2\tau})} \nonumber\\
&=& (\cdots)\sum_{n=0}^{\nu-1}\frac{2\epsilon_{n-\nu}-\frac{1}{\tau}}{\epsilon_n(\omega+\frac{1}{2\tau}+\mathbbm{i}m)(2\epsilon_n+\omega+\frac{1}{\tau})(2\epsilon_{n+\omega}+\bm{Dq}^2)(2\epsilon_{n-\nu}-\bm{Dq}^2)}\nonumber\\
&\equiv&(\cdots)\left( \sum\nolimits^{xxy}_{\text{reg}}+\sum\nolimits^{xxy}_{\text{an}}+\sum\nolimits^{xxy}_{\text{an}*} \right),
\eeq
where the last term is less dominant:
\beq
\sum\nolimits^{xxy}_{\text{an}*}&=&\frac{2(\omega+\frac{1}{2\tau})^2}{m^2(2\omega+\bm{Dq}^2)^2}\sum_{n=0}^{\nu-1}\left[ \frac{1}{(\omega+\frac{1}{\tau})\epsilon_n}-\frac{1}{(\frac{1}{\tau}-\omega-\bm{Dq}^2)(2\epsilon_n+2\omega+\bm{Dq}^2)}\nonumber\right.\\
& &-\left.\frac{1}{(\frac{1}{\tau}+3\omega+\bm{Dq}^2)(2\epsilon_n-2\omega-\bm{Dq}^2)} \right]\nonumber\\
&=& \mathcal{O}(\omega^3),
\eeq
and the remaining two terms are:
\beq
\sum\nolimits^{xxy}_{\text{an}}&=&\frac{\omega+\frac{1}{2\tau}}{m^2(2\omega+\bm{Dq}^2)}\sum_{n=0}^{\nu-1}\left[\frac{1}{(\frac{1}{\tau}+3\omega+\bm{Dq}^2)(2\epsilon_n-2\omega-\bm{Dq}^2)}-\frac{1}{(\frac{1}{\tau}-\omega-\bm{Dq}^2)(2\epsilon_n+2\omega+\bm{Dq}^2)}\right] \nonumber\\
&=& -\frac{\tau^3\omega^2}{\pi Tm^2}\left[ \frac{1}{(4\pi T\tau)^2}\phi''(\frac{1}{2})+\frac{1}{2\tau(2\omega+\bm{Dq}^2)}\frac{1}{4\pi T\tau}\phi'(\frac{1}{2}) \right],
\\
\sum\nolimits^{xxy}_{\text{reg}}&=&\frac{\omega+\frac{1}{2\tau}}{m^2(\omega+\frac{1}{\tau})}\sum_{n=0}^{\nu-1}\frac{2(3\omega+\frac{2}{\tau})}{ (2\epsilon_n+\omega+\frac{1}{\tau})(\frac{1}{\tau}+3\omega+\bm{Dq}^2)(\frac{1}{\tau}-\omega-\bm{Dq}^2)}\nonumber\\
&=& \frac{\tau^3\omega^2}{2\pi Tm^2}\frac{1}{4\pi T\tau}\phi'(\frac{1}{2}+\frac{1}{4\pi T\tau}),
\eeq

Similarly, in $I^{xxy}_{3(2)}$:
\beq
I^{xxy}_{3(2)}(\epsilon_n,\omega,\bm{q}) &=& -(\cdots)\sum_{\epsilon_n} \frac{\theta(-\epsilon_n\epsilon_{n+2\omega}) \theta(\epsilon_n\epsilon_{n+\omega}) \lambda(\epsilon_n,-\epsilon_n,\bm{q})  \lambda(\epsilon_{n+2\omega},-\epsilon_{n+2\omega},\bm{q})}{(\epsilon_n+\epsilon_{n+2\omega})(\omega+\frac{1}{2\tau})(\omega+\frac{1}{\tau})(\epsilon_{n+2\omega}+\frac{1}{2\tau}+\mathbbm{i}m)} \nonumber\\
&=& -(\cdots)\sum_{n=0}^{\nu-1}\frac{(2\epsilon_{n-\nu}-\frac{1}{\tau})(2\epsilon_{n+\omega}+\frac{1}{\tau})}{2\epsilon_n (\epsilon_{n-\nu}-\frac{1}{2\tau}-\mathbbm{i}m)(\omega+\frac{1}{2\tau})(\omega+\frac{1}{\tau}) (2\epsilon_{n+\omega}+\bm{Dq}^2) (2\epsilon_{n-\nu}-\bm{Dq}^2)}\nonumber\\
&\equiv&-(\cdots)\left( \sum\nolimits^{xxy}_{\text{reg}'}+\sum\nolimits^{xxy}_{\text{an}'} \right),
\eeq
we have regular and anomalous contributions:
\beq
\sum\nolimits^{xxy}_{\text{an}'}&=&\frac{(\frac{1}{\tau}-\bm{Dq}^2)^2}{(\omega+\frac{1}{2\tau})(\omega+\frac{1}{\tau}) (2\omega+\bm{Dq}^2)}\sum_{n=0}^{\nu-1}\left(\frac{1}{2\epsilon_n-2\omega-\bm{Dq}^2}-\frac{1}{2\epsilon_n+2\omega+\bm{Dq}^2} \right)\nonumber\\
& &\times \frac{1}{(2\epsilon_{n-\nu}-\frac{1}{\tau}-2\mathbbm{i}m)(2\omega+\frac{1}{\tau}+2\mathbbm{i}m)} \nonumber\\
&=& -\frac{\tau^3\omega^2}{4\pi Tm^2\tau(2\omega+\bm{Dq}^2)} \frac{1}{(4\pi T\tau)^2}\phi''(\frac{1}{2}),\\
\sum\nolimits^{xxy}_{\text{reg}'}&=&\frac{2}{(\omega+\frac{1}{2\tau})(\omega+\frac{1}{\tau}) (2\omega+\frac{1}{\tau}+2\mathbbm{i}m)}\sum_{n=0}^{\nu-1}\left[ -\frac{1}{2\epsilon_n } -(\frac{1}{\tau}-\bm{Dq}^2)\nonumber\right.\\
&\times &\left.\left(\frac{1}{2\epsilon_n-2\omega-\bm{Dq}^2}-\frac{1}{2\epsilon_n+2\omega+\bm{Dq}^2} \right)\left(
\frac{1}{2\epsilon_{n-\nu}-\frac{1}{\tau}-2\mathbbm{i}m}-\frac{1}{2\epsilon_n} \right)  \right]\nonumber\\
&=& -\frac{\tau^3\omega^2}{2\pi Tm^2}\left[\frac{1}{(4\pi T\tau)^2}\phi''(\frac{1}{2})+\frac{1}{4\pi T\tau}\phi'(\frac{1}{2})\right].
\eeq

Thus, in total, the regular MT contributions are:
\beq
Q_{\text{MT(reg)}}^{xxy}&= \kappa^{\text{reg}}_{\text{MT}}e^3 \tau^2\omega^2\int_{|q_x|\leq\xi_x^{-1}}\int_{|q_y|\leq\xi_y^{-1}}\frac{dq_xdq_y}{4\pi^2s}\frac{1}{\sqrt{\epsilon+\xi^2_xq_x^2+\xi^2_yq_y^2}\sqrt{\epsilon+\xi^2_xq_x^2+\xi^2_yq_y^2+r}},
\eeq
where:
\beq
\kappa^{\text{reg}}_{\text{MT}}=\frac{\mathbbm{i}T\tau }{4m^2}\left\langle {v_x^{++}(\bm{p})v_x^{++}(\bm{p})v_y^{+-}(\bm{-p})} \right \rangle_{F.S.}\left[ \frac{5}{4\pi T\tau}\phi'(\frac{1}{2}+\frac{1}{4\pi T\tau})-\frac{2}{4\pi T\tau}\phi'(\frac{1}{2})-\frac{6}{(4\pi T\tau)^2}\phi''(\frac{1}{2}) \right]<0,
\eeq
is a model-dependent non-divergent constant. For $r\to 0$, it gives rise to a sub-leading divergence at $T\to T_c$ and $\epsilon=\ln(T/T_c)\to 0$:
\beq
Q_{MT(reg)}^{xxy}&= \frac{\kappa^{\text{reg}}_{\text{MT}} e^3\tau^2\omega^2}{4\pi\xi_x\xi_yl_z}\ln(1+\frac{1}{\epsilon})\to\chi^{xxy}_{MT(reg)}(\omega_1,\omega_2)=\frac{\kappa^{\text{reg}}_{\text{MT}} e^3\tau^2}{4\pi\xi_x\xi_yl_z}\ln(1+\frac{1}{\epsilon})
\eeq

On the other hand, the anomalous MT contributions are:
\beq
Q_{\text{MT(an)}}^{xxy}&= \kappa^{\text{an}}_{\text{MT}}e^3\tau^2(\eta \tau)^{-1}\omega^2 \int_{|q_x|\leq\xi_x^{-1}}\int_{|q_y|\leq\xi_y^{-1}}\frac{dq_xdq_y}{4\pi^2 l_z}\frac{1}{\sqrt{\epsilon+\xi^2_xq_x^2+\xi^2_yq_y^2}\sqrt{\epsilon+\xi^2_xq_x^2+\xi^2_yq_y^2+r}(\bm{Dq}^2+{2}/{\tau_\phi})},
\eeq
where:
\beq
\kappa^{\text{an}}_{\text{MT}}=\frac{\mathbbm{i}\eta T \tau }{8m^2}\left\langle {v_x^{++}(\bm{p})v_x^{++}(\bm{p})v_y^{+-}(\bm{p})} \right \rangle_{F.S.}\left[\frac{1}{(4\pi T\tau)^2}\phi''(\frac{1}{2}) -\frac{2}{4\pi T\tau}\phi'(\frac{1}{2}) \right]>0,
\eeq
is another model-dependent non-divergent constant. For $r\to 0$,
\beq
Q_{\text{MT(an)}}^{xxy}&= \frac{\kappa^{\text{an}}_{\text{MT}} e^3 \tau^2\omega^2}{8\pi\xi_x\xi_yl_z(\epsilon-\gamma_\phi)}\ln(\frac{\epsilon+\epsilon\gamma_\phi}{\gamma_\phi+\epsilon\gamma_\phi})\to\chi^{xxy}_{\text{MT(an)}}(\omega_1,\omega_2)=\frac{\kappa^{\text{an}}_{\text{MT}} e^3\tau^2}{8\pi\xi_x\xi_yl_z(\epsilon-\gamma_\phi)}\ln(\frac{\epsilon}{\gamma_\phi}),
\eeq
has a similar, sub-leading divergence $\propto \ln(\epsilon)$ for finite $\gamma_\phi$, where $\gamma_\phi={2\tau\eta}/{\tau_\phi}$, $\tau_\phi$ and $\eta$ are discussed in the main text. For the extreme scenario where $\gamma_\phi\to 0$ and $r\ne0$, on the other hand, the anomalous term could, in principle, become as singular as AL contributions.

\subsubsection{The DOS contribution}

The contribution to the nonlinear Hall effect from the DOS fluctuation, shown in Fig. \ref{fig:FDAppendix}(b), is expressed as the following:
\beq
Q_{\text{DOS}}^{xxy}&= e^3 T\sum_{\Omega_k} \int \frac{{\rm d}^3\bm{q}}{(2\pi)^3} \sum_{xxy}(\Omega_k,\omega_1,\omega_2,\bm{q}) L(\Omega_k,\bm{q}),
\eeq
where:
\beq
\sum\nolimits^{xxy}(\Omega_k=0,\omega_1,\omega_2,\bm{q}) &=&T\sum_{\epsilon_n} \lambda(\epsilon_n,-\epsilon_n,\bm{q})^2 I^{xxy}(\epsilon_n,\omega_1,\omega_2,\bm{q}), \\
I^{xxy}(\epsilon_n,\omega_1,\omega_2,\bm{q})&=&\int \frac{{\rm d}^3\bm{p}}{(2\pi)^3} v_x^{++}(\bm{p})v_x^{+-}(\bm{p})v_y^{-+}(\bm{p}) \nonumber\\
&\times&[G^K_+(\epsilon_n,\bm{p})]^2 G^K_+(\epsilon_n+\omega_1,\bm{p})  G^K_-(\epsilon_n+\omega_1+\omega_2,\bm{p}) G^{-K}_+(\Omega_k-\epsilon_n,\bm{q-p})\nonumber\\
&=&-\nu\int_{-\infty}^\infty \frac{d\xi \left\langle {v_x^{++}(\bm{p})v_x^{+-}(\bm{p})v_y^{-+}(\bm{p})} \right \rangle_{F.S.} }{(i\Tilde{\epsilon_n}-\xi)^2[i(\Tilde{\epsilon_n}+\omega_1)-\xi][i(\Tilde{\epsilon_n}+\omega_1+\omega_2)-\xi+2m](i\Tilde{\epsilon_n}+\xi)},
\eeq
where we have suppressed $\bm{q}$ in the Green's functions. After the integration of $\xi$ and setting $\omega=\omega_1=\omega_2$, we have:
\beq
I^{xxy}(\epsilon_n,\omega,\bm{q})
&=&{2\pi\mathbbm{i}\nu}\left\langle {v_x^{++}(\bm{p})v_x^{+-}(\bm{p})v_y^{-+}(\bm{p})} \right \rangle_{F.S.}
\left\{  \frac{{\rm sign}(\epsilon_n)}{(2\Tilde{\epsilon_n})^2(\Tilde{\epsilon_n}+\Tilde{\epsilon_{n+\omega}})(\Tilde{\epsilon_n}+\Tilde{\epsilon_{n+2\omega}}-2\mathbbm{i}m)}\nonumber\right.\\
 &+&\left.\frac{\theta(-\epsilon_n\epsilon_{n+\omega})}{(\omega+\tau^{-1})^2(2\omega+\tau^{-1}-2\mathbbm{i}m)(2\epsilon_n+\omega)} \right\},
\eeq
and then we can expand $\sum_{xxy}$ to the lowest non-vanishing $\omega^2$ order:
\beq
\sum\nolimits^{xxy}(\epsilon_n,\omega,\bm{q})&\equiv&-{\kappa_{\text{DOS}}\nu\tau^2\omega^2}/T=\frac{\mathbbm{i}\nu\tau^3\omega^2}{4 m^2}\left\langle {v_x^{++}(\bm{p})v_x^{+-}(\bm{p})v_y^{-+}(\bm{p})} \right \rangle_{F.S.} \\\nonumber
 &\times&\left\{\frac{1}{4\pi T\tau} [\phi'(\frac{1}{2}+\frac{1}{4\pi T\tau})-\phi'(\frac{1}{2})]+\frac{1}{(4\pi T\tau)^2}[2\phi''(\frac{1}{2}+\frac{1}{4\pi T\tau})-3\phi''(\frac{1}{2})]-\frac{5}{(4\pi T\tau)^3}\phi'''(\frac{1}{2})\right\},
\eeq
where $\kappa_{\text{DOS}}<0$ is a model-dependent non-divergent constant:
\begin{eqnarray}
    \kappa_{\text{DOS}}&=&-\frac{\mathbbm{i}T\tau\left\langle {v_x^{++}(\bm{p})v_x^{+-}(\bm{p})v_y^{-+}(\bm{p})} \right \rangle_{F.S.}}{4 m^2} \nonumber\\
 &\times&\left\{\frac{1}{4\pi T\tau} [\phi'(\frac{1}{2}+\frac{1}{4\pi T\tau})-\phi'(\frac{1}{2})]+\frac{1}{(4\pi T\tau)^2}[2\phi''(\frac{1}{2}+\frac{1}{4\pi T\tau})-3\phi''(\frac{1}{2})]-\frac{5}{(4\pi T\tau)^3}\phi'''(\frac{1}{2})\right\}.
\end{eqnarray}

Thus, we finally have:
\beq
Q^{\text{DOS}}_{xxy}&=& \kappa_{\text{DOS}} \tau^2\omega^2e^3 \int_{|q_x|\leq\xi_x^{-1}}\int_{|q_y|\leq\xi_y^{-1}}\frac{dq_xdq_y}{4\pi^2s}\frac{1}{\sqrt{\epsilon+\xi^2_xq_x^2+\xi^2_yq_y^2}\sqrt{\epsilon+\xi^2_xq_x^2+\xi^2_yq_y^2+r}}
\nonumber\\
&=& \frac{\kappa_{\text{DOS}} e^3\tau^2\omega^2}{4\pi\xi_x\xi_yl_z}\ln \left[\frac{(\epsilon+r)\left(1-\sqrt{\frac{\epsilon}{\epsilon+r}}\right)^2}{(1+\epsilon+r)\left(1-\sqrt{\frac{1+\epsilon}{1+\epsilon+r}}\right)^2}\right],
\eeq
and the corresponding nonlinear conductivity:
\beq
\chi^{xxy}_{\text{DOS}}(\omega_1,\omega_2)&=&-\frac{1}{\omega_1\omega_2}Q^{\text{DOS},R}_{xxy}(\omega_1,\omega_2)\nonumber\\
&=&-\frac{\kappa_{\text{DOS}}  e^3\tau^2}{4\pi\xi_x\xi_yl_z}\ln \left[\frac{(\epsilon+r)\left(1-\sqrt{\frac{\epsilon}{\epsilon+r}}\right)^2}{(1+\epsilon+r)\left(1-\sqrt{\frac{1+\epsilon}{1+\epsilon+r}}\right)^2}\right],
\eeq
diverges as $\ln(\epsilon)$ for $r\rightarrow 0$ and has an opposite sign to the pristine nonlinear Hall response $\chi_0^{xxy}$.

\subsubsection{The AL contributions}

The AL contributions are summarized in Fig. \ref{fig:FDAppendix}(c) and (d):
\beq
Q_{\text{AL(c)}}^{xxy}&= -e^3 T\sum_{\Omega_k} \int \frac{{\rm d}^3\bm{q}}{(2\pi)^3} C_{xx}^{(a)}(\Omega_k,\omega_1,\omega_2,\bm{q}) L(\Omega_k,\bm{q}) L(\Omega_k+\omega_1+\omega_2,\bm{q}) B_y(\Omega_k,\omega_1+\omega_2,\bm{q}),
\eeq
where:
\beq
B_y(\Omega_k,\omega,\bm{q}) &=& T\sum_{\epsilon_n} \lambda(\epsilon_n+\omega,\Omega_k-\epsilon_n,\bm{q}) \lambda(\epsilon_n,\Omega_k-\epsilon_n,\bm{q}) \\\nonumber
& \times &\int \frac{{\rm d}^3\bm{p}}{(2\pi)^3} v_y^{++}(\bm{p}) G^K_+(\epsilon_n+\omega,\bm{p}) G^K_+(\epsilon_n,\bm{p}) G^{-K}_+(\Omega_k-\epsilon_n,\bm{q-p}),
\\
C_{xx}^{(c)}(\Omega_k,\omega_1,\omega_2,\bm{q}) &=& T\sum_{\epsilon_n} \lambda(\epsilon_n+\omega,\Omega_k-\epsilon_n,\bm{q}) \lambda(\epsilon_n,\Omega_k-\epsilon_n,\bm{q}) \\\nonumber
&\times& \int \frac{{\rm d}^3\bm{p}}{(2\pi)^3} v_x^{-+}(\bm{p}) G^K_-(\epsilon_n,\bm{p}) G^K_+(\epsilon_n+\omega_1,\bm{p}) v_x^{++}(\bm{p}) G^K_+(\epsilon_n+\omega_1+\omega_2,\bm{p}) G^{-K}_+(\Omega_k-\epsilon_n,\bm{q-p}).
\eeq

Since we only care about the physics in the vicinity of the critical point, $B_y$ and $C_{xx}$ are mostly independent of the frequencies $\Omega_k$ and $\omega$, which we simply set to 0. Next, we replace the integral over momentum with the integral over $\xi$ and the density of states $\nu$:
\beq
B_y(\bm{q})&\equiv& B_y(0,0,\bm{q})=-T\sum_{\epsilon_n} \lambda(\epsilon_n,-\epsilon_n,\bm{q})^2 \int_{-\infty}^\infty \frac{\nu d\xi}{(i\Tilde{\epsilon_n}-\xi)^2} \left\langle \frac{v_y^{++}(\bm{p})}{i\Tilde{\epsilon_n}+\xi-\bm{v_p\cdot q}} \right \rangle_{F.S.},
\\
C_{xx}^{(c)}(\omega_1,\omega_2,\bm{q}) &\equiv& C_{xx}^{(c)}(0,\omega_1,\omega_2,\bm{q}) = -T\sum_{\epsilon_n}
 \lambda(\epsilon_n,-\epsilon_n,\bm{q})^2 \nonumber\\
&\times& \int_{-\infty}^\infty \frac{\nu d\xi}{(i\Tilde{\epsilon_n}-\xi+2m)[i(\Tilde{\epsilon_n}+\omega_1)-\xi][i(\Tilde{\epsilon_n}+\omega_1+\omega_2)-\xi]} \left\langle \frac{v_x^{-+}(\bm{p})v_x^{++}(\bm{p})}{i\Tilde{\epsilon_n}+\xi-\bm{v_p\cdot q}} \right \rangle_{F.S.}.
\eeq

In addition, the major contribution within the integral is concentrated in the region of small $\bm{q}$ due to the large correlation length. Therefore, we expand $B_y$ and $C_{xx}$ around $\bm{q}=0$ and keep its leading order:
\beq
B_y(\bm{q})&=&-\pi T\nu \sum_{n=0}^\infty \frac{\left\langle {v_y^{++}(\bm{p})^2}\right \rangle_{F.S.}}{(\epsilon_n+\frac{1}{2\tau})\epsilon_n^2} q_y\nonumber\\
& = & 2\nu \tau^2 \left[ \phi(\frac{1}{2}+\frac{1}{4\pi T\tau}) - \phi(\frac{1}{2})-\frac{1}{4\pi T\tau}\phi'(\frac{1}{2}) \right] \left\langle {v_y^{++}(\bm{p})^2}\right \rangle_{F.S.} q_y,
\\
C_{xx}^{(c),R}(\omega_1,\omega_2,\bm{q}) &=&-\pi T\nu \tau^2(2\omega_1+\omega_2) \sum_{n=0}^\infty \frac{\left\langle {v_x^{++}(\bm{p}) v_x^{-+}(\bm{p}) v_y^{++}(\bm{p})}\right \rangle_{F.S.}}{4m^2(\epsilon_n+\frac{1}{2\tau})\epsilon_n^2} q_y.
\eeq

Similarly, for the contribution of Fig. \ref{fig:FDAppendix}(d), we have:
\beq
C_{xx}^{(d),R}(\omega_1,\omega_2,\bm{q}) =\pi T\nu \tau^2(\omega_1+2\omega_2) \sum_{n=0}^\infty \frac{\left\langle {v_x^{++}(\bm{p}) v_x^{+-}(\bm{p}) v_y^{++}(\bm{p})}\right \rangle_{F.S.}}{4m^2(\epsilon_n+\frac{1}{2\tau})\epsilon_n^2} q_y.
\eeq

In total, we have:
\beq
C_{xx}^R(\omega_1,\omega_2,\bm{q})&\equiv& C_{xx}^{(a)}(\omega_1,\omega_2,\bm{q}) + C_{xx}^{(b)}(\omega_1,\omega_2,\bm{q})\nonumber\\
&= &-\frac{3\nu\tau^2 (\omega_1+\omega_2)}{m^2}  \left[ \phi(\frac{1}{2}+\frac{1}{4\pi T\tau}) -\phi(\frac{1}{2})-\frac{1}{4\pi T\tau}\phi'(\frac{1}{2}) \right] \left \langle v_x^{++}(\bm{p}) v_x^{-+}(\bm{p}) v_y^{++}(\bm{p})\right \rangle_{F.S.} q_y,
\eeq
and the AL contributions become:
\beq
Q_{\text{AL,R}}^{xxy}(\omega_1,\omega_2)&=& e^3 \int \frac{{\rm d}^3\bm{q}}{(2\pi)^3}6(\omega_1+\omega_2)(2\nu\eta\tau^2)^2 \left \langle v_x^{++}(\bm{p}) v_x^{-+}(\bm{p}) v_y^{++}(\bm{p})\right \rangle_{F.S.}\left\langle {v_y^{++}(\bm{p})^2}\right \rangle_{F.S.} q_y^2\nonumber\\
&\times& \frac{\mathbbm{i}(\omega_1+\omega_2)}{4\pi Tm^2}\int_{-\infty}^{\infty}\frac{dz}{[{\rm sinh(\frac{z}{2T})}]^2}\left[ {\rm Im}L(-\mathbbm{i}z,\bm{q}) \right]^2.
\eeq

Finally, after the integration, we obtain the nonlinear conductivity:
\beq
\chi^{xxy}_{\text{AL}}(\omega_1,\omega_2)&=&-\frac{1}{\omega_1\omega_2}Q_{\text{AL,R}}^{xxy}(\omega_1,\omega_2)\nonumber\\
&=&\frac{e^3\mathbbm{i}}{\pi l_z m^2}\frac{\left \langle v_x^{++}(\bm{p}) v_x^{+-}(\bm{p}) v_y^{++}(\bm{p})\right \rangle_{F.S.}}{\sqrt{\epsilon(\epsilon+r)}\sqrt{\left\langle {v_x^{++}(\bm{p})^2}\right \rangle_{F.S.}\left\langle {v_y^{++}(\bm{p})^2}\right \rangle_{F.S.}}}.
\eeq
which has the same sign as $\chi^{xxy}_{0}$ and is a leading order contribution for $T\to T_c$.
\\

\section{Additional parameter settings and results of the tilted massive Dirac model}

The condition of the Fermi surface is straightforward from Eq. \ref{eq:hamk} in the main text:
\begin{eqnarray}
    (E_F-tk_x)^2 = v^2(k_x^2+k_y^2)+m^2,
\end{eqnarray}
for the $K$ valley, with the parameters satisfying:
\begin{eqnarray}
    v>t,~~~E_F>\frac{\sqrt{v^2-t^2}}{v}m,
\end{eqnarray}
for a closed Fermi surface. Hence, the resulting Fermi surface is an ellipse with major axis and minor axis as:
\begin{eqnarray}
    R_x=\sqrt{\frac{v^2E_F^2}{(v^2-t^2)^2}-\frac{m^2}{v^2-t^2}},~~~R_y=\sqrt{\frac{E_F^2}{v^2-t^2}-\frac{m^2}{v^2}}.
\end{eqnarray}

In particular, $t$ controls the inclination of the Dirac cone, breaks the inversion symmetry, and increases the Fermi-surface anisotropy; $v$ tunes the Fermi velocity (steepness) of the Dirac cone and affects the distribution of the Berry curvature; the mass $m$ opens a gap in the energy spectrum, influencing the Berry curvature concentration and smoothness; lifting  $E_F$ enlarges the Fermi surface, spreading the Berry curvature over a broader range in the momentum space. Collectively, these model parameters influence the Berry curvature distribution and the shape and size of the Fermi surface, which impact the validity of the approximations implicit in the semiclassical picture.

\begin{figure}
\centering
\includegraphics[width=16cm]{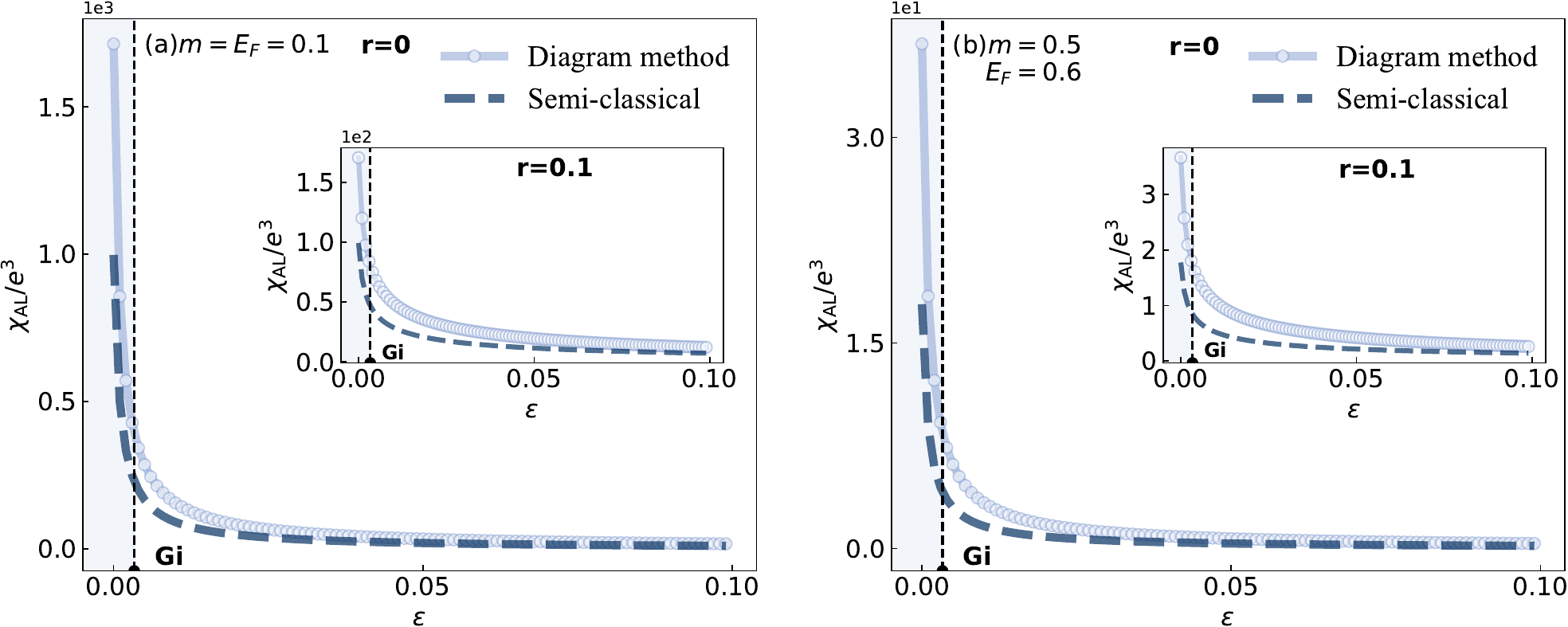}
\caption{For (a) $m=0.1$, $E_F=0.1$ and (b) $m=0.5$, $E_F=0.6$, either the small mass leads to a concentrated Berry curvature distribution or the Fermi energy is close to the band bottom leading to a small pocket, causing potential deviations in approximations behind the semiclassical theory of the nonlinear Hall response in Eq. \ref{eq:Semi-AL} in the main text. Nevertheless, comparisons with the real parts of the AL contributions $\chi_{\text{AL}}$ from Eq. \ref{eq:ALcontr} in the main text show semi-quantitative consistency, both in the 2D limit $r\to 0$ and with finite inter-layer coupling $r$ in quasi-2D (insets). Here, we set $t=0.04$ and $v=0.3$. } \label{fig:semi_appendix}.
\end{figure}

In addition to the examples in Fig. \ref{fig:chi} in the main text, we show in Fig. \ref{fig:semi_appendix} additional settings where we compare the semiclassical and quantum approaches for the Cooper pair contributions, Eqs. \ref{eq:Semi-AL} and \ref{eq:ALcontr} in the main text. In particular, here, we investigate scenarios challenging the validity of the approximations underlining the semiclassical picture, as we have discussed in the main text. Indeed, compared with the quantum results, the semiclassical results exhibit observable discrepancies yet still offer more or less semi-quantitative guidelines.

\end{widetext}

\end{document}